\documentclass[a4paper]{aastex63}
\usepackage{hyperref}
\usepackage{cleveref}
\graphicspath{ {./} }

\begin{document}

\title{Intrinsic Color Indices of Early-Type Dwarf Stars}

\author{Dingshan Deng}
\affiliation{Department of Astronomy,
  Beijing Normal University,
  Beijing 100875, China}

\author{Yang Sun}
\affiliation{Department of Astronomy,
  Beijing Normal University,
  Beijing 100875, China}

\author{Mingjie Jian} 
\affiliation{Department of Astronomy,
 School of Science,
 The University of Tokyo,
 7-3-1 Hongo, Bunkyo-ku, Tokyo 113-0033, Japan}
\email{mingjie@astron.s.u-tokyo.ac.jp}

\author{Biwei Jiang}
\affiliation{Department of Astronomy,
  Beijing Normal University,
  Beijing 100875, China}

\author{Haibo Yuan}
\affiliation{Department of Astronomy,
  Beijing Normal University,
  Beijing 100875, China}

\begin{abstract}

Early-type stars are short lived and scarce in comparison with other types. Based on the recently released catalogs of early type stars from the largest LAMOST spectroscopic survey, the intrinsic colors of the stars with effective temperature up to 32,000\,K are determined for the bands from ultraviolet to infrared by using the blue-edge method. Analytic relations are derived for the intrinsic color index with the effective temperature for the \emph{WISE}, 2MASS, \emph{Gaia}, APASS, SDSS, Pan-STARRS1, and \emph{GALEX} bands. The results are generally consistent with previous works. In addition, the intrinsic colors of O-type dwarfs and OB supergiants are roughly estimated.


\end{abstract}



\section{Introduction} \label{sec:intro}

Intrinsic color index, or simply called intrinsic color, is a basic stellar parameter related to stellar properties such as atmospheric temperature, metallicity etc.
It can be adopted as a reference in astronomical studies such as in Hertzsprung-Russell diagrams and used to analyze spectral energy distributions (SED; \citealt{the_spectral_1986}; \citealt{lee_photometry_1970}).
Meanwhile, the intrinsic colors are the key to estimate interstellar reddening  and extinction which are important probes of interstellar dust.

Early-type stars, referring to O-, B- and A-type stars in this work, are very hot and massive stars so that they are short-lived.
Mostly located in the spiral arms, or other dusty regions from birth, they are usually immersed in dense interstellar clouds.
Thus, the number of observable early-type stars are highly limited compared with that of later types.
Besides, early-type stars are famous for their strong stellar wind to produce considerable circumstellar material \citep{garcia-segura_dynamical_1996}, which brings about uncertainty in determining their stellar properties.
Consequently, their intrinsic colors are not well determined.

The intrinsic colors of early-type stars were studied previously by many works in an empirical approach, for example, \citet{johnson_astronomical_1966}, \citet{fitzgerald_intrinsic_1970}, \citet{kuriliene_intrinsic_1977}, \citet{whittet_infrared_1980}, \citet{straizys_intrinsic_1987}, and \citet{wegner_extinction_1993, wegner_intrinsic_1994, wegner_intrinsic_2014}.
Although different in details, the principle method adopts two-color diagrams and specific de-reddening laws.
They used the experimental formula $V- \lambda =a(B-V)+b$, which depends on the assumption that two colors show a linear relationship in some particular range of stellar parameters.
Once an intrinsic color index (e.g. $(B-V)_0$) is determined, the other intrinsic colors can be calculated by substituting it into the formula as $(V- \lambda)_0=a(B-V)_0+b$, where the constants $a$ and $b$ are derived from fitting the observed colors.
There are a few problems with this method.
Firstly, the linear relation only appears in very limited ranges of stellar parameters and some colors.
Figure 2 of \citet{chen_three-dimensional_2014} displayed the relation of various intrinsic color indexes with $(g-i)_0$, which clearly showed the variety of the relations even including inverse correlation in some cases, and almost in no case where all the color indexes are linearly related.
Secondly, they need to make their own assumptions or adopt an established de-reddening law to determine the first intrinsic color $(B-V)_0$, which would reduce the accuracy.
\citet{johnson_astronomical_1966} assumed that the nearest stars within 100\,pc from the Sun were not affected by interstellar extinction and they took the mean value of the observed colors of these stars as intrinsic.
In the work of \citet{wegner_extinction_1993, wegner_intrinsic_1994, wegner_intrinsic_2014}, he adopted the PWK de-reddening law from \citet{papaj_intrinsic_1990} while \citet{papaj_intrinsic_1993} found that adopting their de-reddening laws brought about some degree of uncertainty.
Thirdly, this procedure requires the extinction law to be the same for all the stars in different sightlines or interstellar environments, which may only be valid in some particular regions such as in the OB associations \citep{krelowski_extinction_1987}.
Additionally, the accuracy of intrinsic color is severely affected by the reference color index ($(B-V)_0$ here), i.e., the intrinsic colors are not derived independently.


Meanwhile, synthetic photometry of stellar atmosphere models were utilized to determine the intrinsic colors for early-type stars by other works. In such approach, the effective temperatures (\(T_{\mathrm{eff}}\)) are determined from modeling the spectral features, and the SED of these models can then be convolved with the appropriate sensitivity functions to calculate the intrinsic colors.
As an early example, \citet{flower_bolometric_1975} determined the relationship between Teff and (B-V)o for the late-type (M-type) giants and supergiants. Combining with other results (such as \citealt{code_empirical_1976} for stars with \(T_{\mathrm{eff}}\) $>$ 11000\,K), intrinsic colors for supergiants, giants and dwarfs were presented in \citet{flower_transformations_1977}.
Then, by the aid of ATLAS (\citealt{kurucz_model_1979}) model, \citet{bessell_model_1998} determined the intrinsic colors for early-type stars (\(T_{\mathrm{eff}}\) $>$ 9000\,K) within the Johnson-Cousins-Glass UBVJHK system, and \citet{martins_ubvjhk_2006} performed a similar calculation for O-type stars. In recent years, the PARSEC model \citep{bressan_parsec:_2012} is popularly used to calculate stellar intrinsic colors for its updated parameters, flexibility as well as convenient availability online, which is taken into comparison with the present work.

Recently, the so-called blue-edge method is developed to determine stellar intrinsic colors and applied to the stars of all except early spectral types.
The idea is originated from \citet{ducati_intrinsic_2001}, which assumed the bluest ones in a sample of stars at a given spectral type (and luminosity class) suffer no interstellar extinction and their observed colors are the same as the intrinsic color of this spectral type.
This assumption is very reasonable as far as the sample is large enough and includes the zero-reddening stars.
This method avoids the shortcomings aforementioned with no assumption on interstellar extinction law or linear relationship between intrinsic colors.
Originally, \citet{ducati_intrinsic_2001} used a catalog of 3946 sources for dwarfs, giants and subgiants due to the limit of observational data and could only obtain a very rough estimation of stellar intrinsic colors according to spectral types.
With the development of both spectroscopic and photometric observation, significantly larger datasets are collected and this method is developed. With the stellar parameters from the LAMOST and APOGEE spectroscopic surveys, \citet{wang_universality_2014}, \citet{xue_precise_2016}, \citet{jian_revision_2017} and \citet{sun_ultraviolet_2018} derived the analytic relations of stellar intrinsic colors with \(T_{\mathrm{eff}}\) in the infrared and ultraviolet bands from the blue-edge in the observed color versus \(T_{\mathrm{eff}}\) diagram, which were used to determine the interstellar extinction law.
The accuracy of the blue-edge method depends mainly on the photometric uncertainty (which is unavoidable in any method) and the uncertainty of stellar parameters. The accuracy is generally very high since those two uncertainties are usually small.

However, the early-type stars are not included in these studies of stellar intrinsic colors by the blue-edge method. The key reason lies on the lack of appropriate sample of early-type stars.
Several works have been done to identify OB stars.
The Catalog of Galactic OB stars \citep{reed_catalog_2003} collects nearly 16,200 OB stars, including 10,669 OB stars from 12,235  Case-Hamburg Galactic plane luminous-stars and 5,500 additional stars from other literatures. However, over 8,000 of them have no information of spectral classification. Besides, the spectral classification is not based on  homogeneous criteria since this is an assembly of various catalogs.
The catalog by \citet{gontcharov_ob_2008, gontcharov_spatial_2012} contains 20,514 OV-A0V stars from the Tycho-2 catalog. Because they selected early-type stars from color-color diagrams, no further spectral classification is provided. There must be some contamination of other types of stars as photometry cannot determine the spectral type as convincing as spectroscopy.  According to the tests with Hipparcos Input Catalogue (HIC, \citealt{turon_version_1993}) and Tycho-2 Spectral Types (TST) catalog \citep{wright_tycho-2_2003}, 14,732 (94 \%) of them are of O- to A0-type \citep{gontcharov_spatial_2012}.
\citet{mohr-smith_deep_2017} selected 14,900 early-type stars with 5,915 high confidence O-B2 stars with \(T_{\mathrm{eff}}\) from SED fitting, but these stars are limited in 42 deg$^2$ in the Carina Arm region with high interstellar extinction.
Galactic O-Star Spectroscopic Survey (GOSSS; \citealt{maiz_apellaniz_galactic_2011}; \citealt{sota_galactic_2014}; \citealt{maiz_apellaniz_galactic_2016}) is a project dedicated to O-type stars.
With high signal to noise ratio (S/N) and the resolution of $\sim$2500, this survey yields more than 1000 Galactic O stars, which should be the biggest catalog for O stars with spectral classification, though the number of O stars is still limited.

With specific design, the Large Sky Area Multi-Object Fiber Spectroscopic Telescope (LAMOST) has obtained over 10 million stellar spectra.
This huge database provides a possibility to expand the catalog of early-type stars, in particular OB stars.
Although identifying OB stars is not very difficult, deriving the stellar parameters is hard due to the scarcity of spectral lines and the fact that LAMOST survey makes no absolute flux calibration.
\cite{liu_catalog_2019} identified about 16,000 OB stars from the data release 5 (DR5) of LAMOST survey \footnote{http://dr5.lamost.org/}, ever the largest reliable catalog of the OB stars.
Furthermore, Liu et al. (in prep) determined the basic stellar parameters, i.e. \(T_{\mathrm{eff}}\) and \(\log\,g\), for about 9000 stars in this LAMOST OB star catalog.
This newly updated OB star catalog brings us a chance to analyze their intrinsic colors with a much larger sample by the blue-edge method. For the A-type stars with  \(T_{\mathrm{eff}}\) $<$ 8000\,K whose stellar parameters were derived by the LAMOST pipeline, \citet{jian_revision_2017} already determined their intrinsic colors by the blue-edge method.
In order to expand the spectral range coverage, the A-type stars with \(T_{\mathrm{eff}}\) $>$ 8000\,K from the LAMOST DR5 value-added catalog \citep{xiang_lamost_2017} are included in this study.

Besides the development of spectroscopic surveys, a number of new photometric surveys are conducted.
The most widely used includes the \emph{WISE}, 2MASS, \emph{Gaia}, APASS, SDSS, Pan-STARRS1, and \emph{GALEX} surveys.
These surveys use the filters previously undefined, correspondingly the related intrinsic color indexes need to be determined for the early-type stars.

In this work, we determine the intrinsic colors of early-type stars with recently released large catalogs from the LAMOST spectroscopy survey using the blue-edge method for some traditional and innovatory filters.
We describe the data in Section~\ref{sec:Data} and the details of the blue-edge method in Section~\ref{sec:Method}.
The result and discussion will be presented in Section~\ref{sec:Results}, and summary in Section~\ref{sec:Summary}.

\section{Data}
\label{sec:Data}

Photometric and spectroscopic data are from several large-scale surveys, which gives us the opportunity to collect sufficient data of early-type stars needed for  more precise determination of their intrinsic colors.

\subsection{Spectroscopic Data}

LAMOST (Large Sky Area Multi-Object Fiber Spectroscopy Telescope; \citealt{cui_large_2012}) is a reflective Schmidt telescope  at the Xinglong Station of the National Astronomical Observatory of China.
We used two catalogs from the LAMOST survey,  one of the A-type stars from \citet{xiang_lamost_2017} and the other of the OB stars from \citet{liu_catalog_2019}. They  both contain  the basic stellar parameters: effective temperature \(T_{\mathrm{eff}}\) and surface gravity \(\log\,g\).
The A-type catalog is selected from the LAMOST DR5 value-added catalog \citep{xiang_lamost_2017} with \(T_{\mathrm{eff}}\)  from 8000\,K to 10000\,K including 205,068 high-temperature A-type stars.
We checked the identification of A-type stars of this catalog. First, the catalog is cross identified with the LAMOST DR5 database which classifies the stars by template match. It is found that 1617 stars are classified as white dwarfs, 98 as quasars and 995 as galaxies in the LAMOST DR5 database. Moreover, some objects are labeled as 'unknown' due to the low S/N spectrum, and some objects are not included in the DR5 database. Additional 188 white dwarfs are recognized by their faintness in \emph{Gaia}/$G$  ($G > 15$) and close distance with \emph{Gaia} distance $< 400$\,pc which bring them together in an isolated area in the \emph{Gaia}/distance-\emph{Gaia}/$G$ diagram. Altogether, 2898 objects are removed from the A-type catalog of \citet{xiang_lamost_2017} for mis-classification.
The OB star catalog is also based on the DR5 but resulted from a dedicated study of the spectral line indices, containing 8582 OB stars \citep{liu_catalog_2019}, with \(T_{\mathrm{eff}}\) ranging within 10000-32000\,K (Liu et al. in prep).
In addition to \(T_{\mathrm{eff}}\) and \(\log\,g\), the stellar metallicity [Fe/H] and its error are available as well for the A-type stars, while not for the OB stars.

\subsection{Photometric Data}


In the optical, we use the data from the \emph{Gaia}, Pan-STARRS1, APASS and SDSS surveys. \emph{Gaia} (Gaia Space Telescope; \citealt{collaboration_gaia_2016}) performed all-sky photometric observations in 3 passbands: $G$ ($\lambda_{\rm eff}$=6730 \AA), \(G_{\mathrm{BP}}\) (5320 \AA) and \(G_{\mathrm{RP}}\) (7970 \AA) with very high accuracy through wide dynamical range benefitted from the space observation.
In \emph{Gaia} DR2 \citep{collaboration_gaia_2018}, its precision in $G$  is about 1 mmag at the bright ($G<13$) end to around 20 mmag at the faint end ($G = 20$); the precisions in other two bands are a few mmag at the bright end to around 200 mmag at the faint end.
We make no use of $G$ band data because the bandwidth is so large that the effective wavelength would be very sensitive to spectral energy distribution.
Pan-STARRS (Panoramic Survey Telescope and Rapid Response System; \citealt{chambers_pan-starrs1_2016}) conducted by a 1.8m telescope  in Hawaii surveys in five passbands: $g$ ($\lambda_{\rm eff}$=4866 \AA), $r$ (6215 \AA), $i$ (7545 \AA), $z$ (8679 \AA) and $y$ (9633 \AA), which extends the wavelength to infrared adjacent to the common JHK bands.
The newly released DR2 is used with a saturation magnitudes at about 12-14\,mag.
Since the Johnson system is classical in photometry, it would be convenient to compare with previous works by adding the UBV photometric results. For the $B$ and $V$ bands, the APASS DR9 data \citep{henden_apass_2014, henden_vizier_2016} is supplemented, which is reliable from about 10.0 mag to 17.0 mag in V-band. Similar to $B-V$, $U-B$ is a widely analyzed color, and more sensitive to \(T_{\mathrm{eff}}\) for early-type stars.
However, the U band is not included in recent photometric surveys so that the data is not sufficient. As a substitution, $(u'-g')_0$ from SDSS DR12 \citep{york_sloan_2000, alam_eleventh_2015} is analyzed since the SDSS/$u'$ ($\lambda_{\rm eff}$=3596 \AA) and $g'$ (4639 \AA) filters are similar to the APASS/$U$ (3663 \AA) and $B$ (4361 \AA) \citep{bessell_standard_2005}, though $(u'-g')_0$ may be bigger than $U-B$ due to slightly wider wavelength difference between the two filters.

In the infrared, we use the data from the 2MASS and \emph{WISE} surveys.
2MASS \citep{skrutskie_two_2006} is a near-infrared all-sky survey finished in 2001.
The photometric limits in the observed $J$ (\(\lambda_{\mathrm{eff}} = \)1.25 \micron), $H$ (1.65 \micron) and $K_S$ (2.17 \micron) bands are 15.8, 15.1 and 14.3\,mag at SNR of 5 respectively.
The \emph{WISE} is an infrared space telescope launched by NASA in December 2009 \citep{wright_wide-field_2010}.
It surveyed in four bands, $W1$ (\(\lambda_{\mathrm{eff}} = \)3.35 \micron), $W2$ (4.60 \micron), $W3$ (11.56 \micron) and $W4$ (22.08 \micron) with a bandwidth of 0.66, 1.04, 5.51, and 4.10 \micron, respectively.
After \emph{WISE} All-Sky \citep{cutri_vizier_2013} survey, the NEOWISE warm mission \citep{mainzer_preliminary_2011} only observed in $W1$ and $W2$.
Consequently, the data in the $W3$ and $W4$ bands are fewer than other bands, namely 7628 stars for $W3$ and 1151 stars for $W4$ in our sample.
Moreover, the photometric quality in $W4$ is relatively poor, which is not used in further analysis.
Fortunately, the spectral energy distribution at wavelength as long as $W4$ should be well approximated by the Rayleigh-Jeans law for hot stars, which implies a rather constant color index in mid- and far-infrared.

For the ultraviolet bands in which the early type stars are bright, we use the data from \emph{GALEX} (Galaxy Evolution Explorer) DR6/7 \citep{martin_galaxy_2005} , which is presently the largest ultraviolet survey.
The photometry of \emph{GALEX} includes two bands: NUV (1750-2800 \AA) and FUV (1350-1740 \AA) with a limiting\,magnitude of 21\,mag and 20\,mag respectively.

The cross identification between catalogs are performed within a radius of 3\arcsec, which is about three times of positional uncertainties, although some catalogs like \emph{Gaia} have higher positional accuracy. The identification should be right when no blending occurs.
In cases there are more than one object within this 3\arcsec \, region centered at the spectroscopic star, the closest one in photometric data is chosen.



\subsection{Data quality control}

We control the data quality for a precise determination of the intrinsic colors by a compromise between the size and the precision of the sample.
Since every catalog has its own claimed errors, our criteria to filter the data coincide with these errors.
For the photometric catalogs, the data are cut within the error of 0.01\,mag, 0.05\,mag, 0.02\,mag, 0.03\,mag, 0.03\,mag, 0.05\,mag, 0.1\,mag, and 0.05\,mag respectively for \emph{Gaia}, SDSS/$u'$\&$g'$, Pan-STARRS (PS1), 2MASS, \emph{WISE}/$W1$\&$W2$, \emph{WISE}/$W3$, \emph{GALEX}, APASS in order.
In addition, the saturation magnitude is set to be 14 for the PS1 and SDSS photometric results.  The quality flag in PS1 is required to be 16 representative of good-quality in the stack; the quality flags `Q' and `q\_mode' in SDSS are required to be 3 and `+', respectively, for selecting the most reliable data.

\begin{figure}
\gridline{\fig{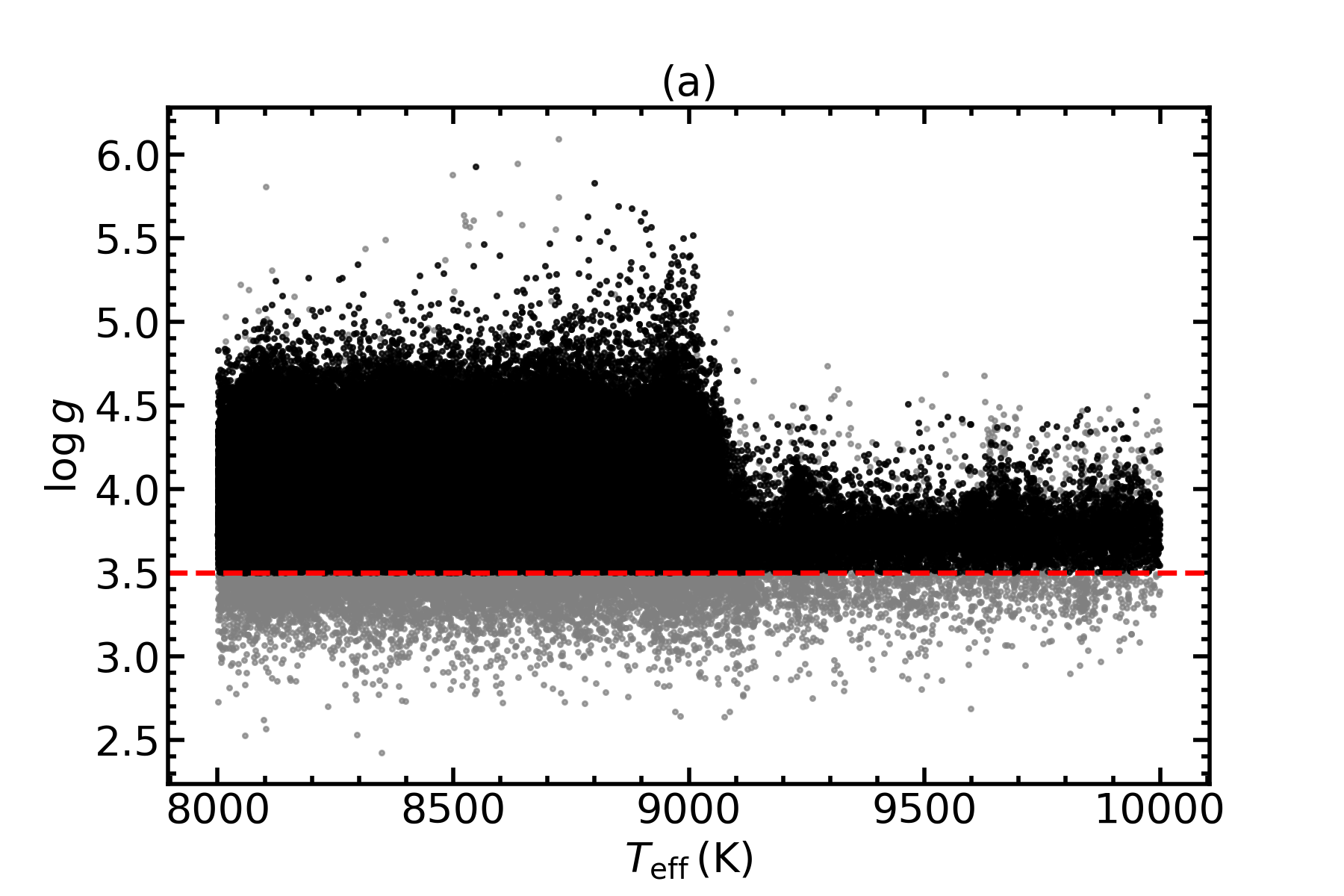}{0.5\textwidth}{}
          \fig{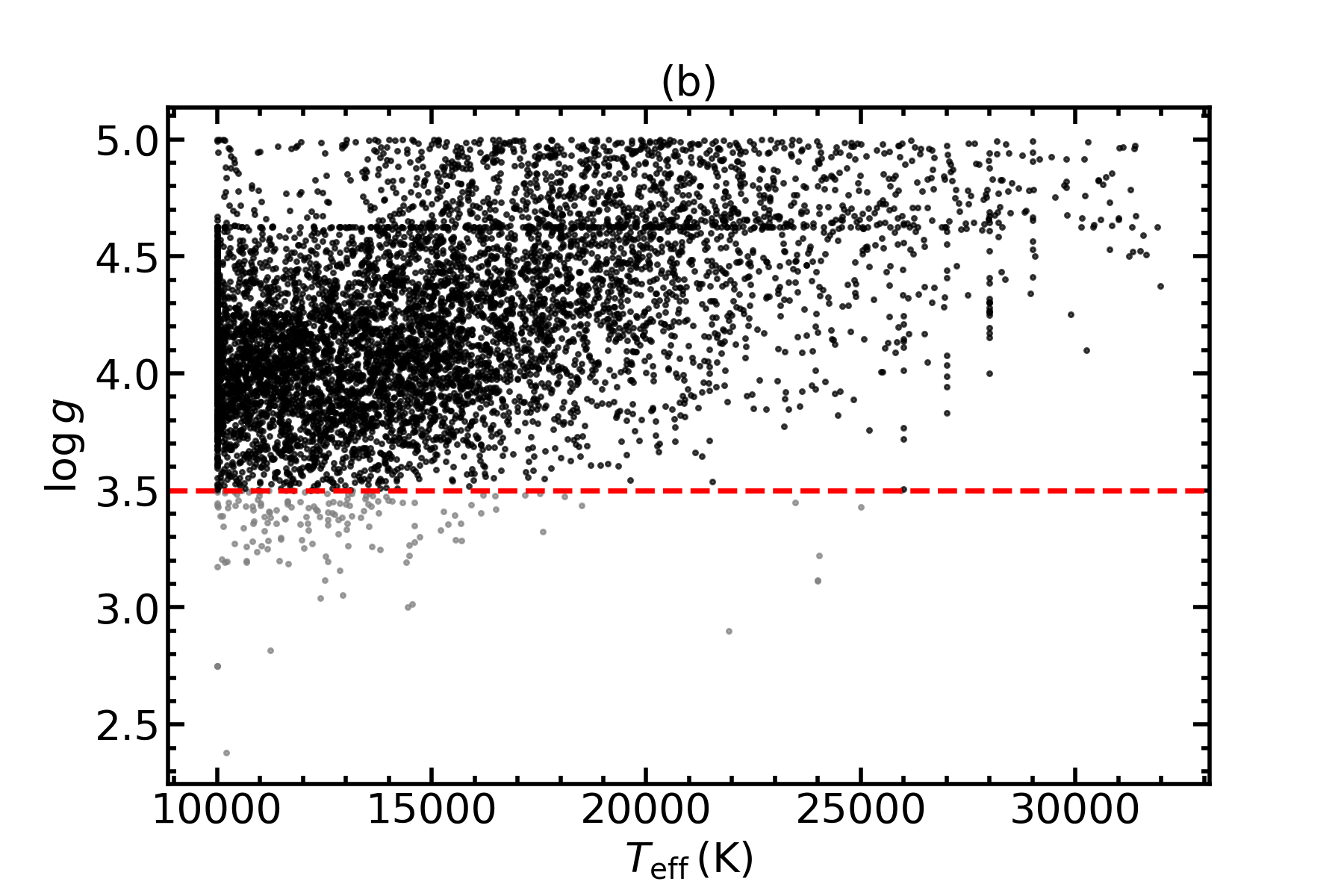}{0.5\textwidth}{}
          }
\caption{Kiel Diagrams for (a) the A-type star and (b) the OB star catalogs. The red line refers to the limit for \(\log\,g\). The grey points denote the original sample and the black points the selected sample according to the criteria of \(\log\,g\) and [Fe/H].}
\label{fig:KielDiagram}
\end{figure}

\begin{figure}
    \centering
    \includegraphics[width=0.9\textwidth]{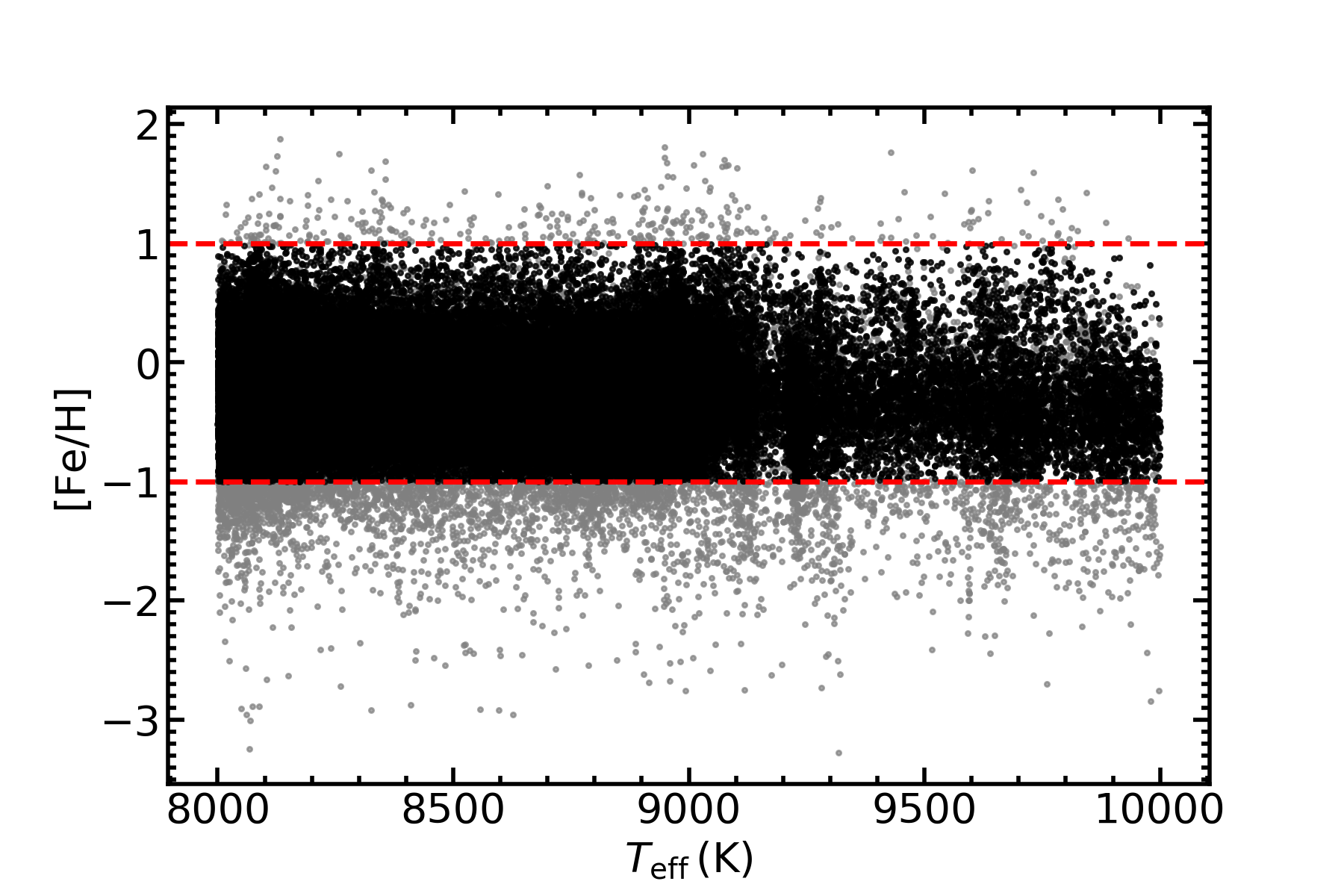}
    \caption{[Fe/H] distribution of the stars from the A-type star catalog. The red dash lines mark our criteria of $-1 < \mathrm{[Fe/H]} < 1$.  The grey points denote the original sample and the black points the selected sample according to the criteria of \(\log\,g\) and [Fe/H].
    \label{fig:FeHdistributionforA}}
\end{figure}

For spectroscopic data from LAMOST, quality control is put on \(T_{\mathrm{eff}}\) and \(\log\,g\).
For A-type star catalog, the relative error of temperature  $\sigma_{T_{\mathrm{eff}}}/T_{\mathrm{eff}}$ is required to be less than 10\%.
While for OB stars, no further quality control on effective temperature is adopted because the error of $T_{\mathrm{eff}}$ is unavailable.
Meanwhile, \(\log\,g\) is constrained to $>3.5$ to pick up the dwarf stars for both the catalogs by taking account of the Kiel Diagram (Figure~\ref{fig:KielDiagram}), following the suggestion by \citet{worley_ambre_2016}.
The giant stars are not studied because their sample is too small to suit  the blue-edge method.
On the metallicity, it has little effect on the intrinsic colors in infrared  \citep{bessell_jhklm_1988}.
In the optical, the metallicity effect starts to appear, while it becomes large for the UV bands \citep{sun_ultraviolet_2018}.
Unfortunately, the catalog for OB stars (\(T_{\mathrm{eff}}\) $> 10000$\,K) provides no measurement of stellar metallicity.
Nevertheless, OB stars are generally young in the Galactic thin disk and thus metal-rich like the Sun. It would be reasonable to assume that the OB stars share the same metallicity with the Sun i.e. $\mathrm{[Fe/H]} = 0$.
Accordingly, a criteria of $-1 < \mathrm{[Fe/H]} < 1$ is set for A-type stars (Figure~\ref{fig:FeHdistributionforA}) and not further divided.

The final numbers of stars in each sample to determine the color indexes are listed in  Table~\ref{tab:samplenumber}, and they are decoded by black dots in Figure~\ref{fig:KielDiagram} where the grey dots decode the stars dropped in further analysis.

\section{The Blue Edge Method}
\label{sec:Method}

Following the method of \citet{wang_universality_2014}, \citet{xue_precise_2016}, \citet{jian_revision_2017}, and \citet{sun_ultraviolet_2018}, we make use of the blue-edge method to determine intrinsic colors of early-type stars from the above selected sample. The key step is to find the blue edge in the \(T_{\mathrm{eff}}\)-color diagrams for the intrinsic color index. Although this method seems to be mature after a series of applications, the application on early-type stars present some new problems. The details of our procedure are following:
\begin{enumerate}
\item Divide the sample into some bins according to \(T_{\mathrm{eff}}\) in the \(T_{\mathrm{eff}}\)-color diagram. The number of sources are very different at different temperature due to both observational bias and intrinsic non-homogenous distribution (the number of early-type stars is less than that of late-type stars), thus we take a variable bin size. For \(T_{\mathrm{eff}}\)  from 8000 to 10000\,K (i.e. A-type stars) where a great number of stars (about $10^5$ sources for each color index) are observed, a size of 50\,K is adopted. While for the hotter (i.e. O- and B-type) stars, the sample is significantly smaller. A bin size of 500\,K and 5000\,K  is adopted for \(T_{\mathrm{eff}}\) $\in$ (10000\,K, 16000\,K) and  \(T_{\mathrm{eff}}\) $>$16000\,K respectively. In addition, a sliding window with a step of 1000\,K is used to compensate for the scarcity of bin numbers after 16000\,K. Although we are compelled to adopt such big bin size, there should be little effect because intrinsic color indices change slowly with \(T_{\mathrm{eff}}\) at the high \(T_{\mathrm{eff}}\) end.  As Table~\ref{tab:samplenumber} shows, the cross-identified sample  between \emph{GALEX}/NUV and APASS/\emph{B} has only 2669 sources, for which  a special bin size of 2000\,K from 10000\,K to 16000\,K is used to include more stars in an individual bin.
  \item Discard the bins with $<10$ stars because the small number would lead to very uncertain definition of the blue edge. For $u'-g'$, those bins at \(T_{\mathrm{eff}}\) $\in [8000, 10000]$\,K with $<50$ stars are discarded to match the photometric quality of SDSS/$u'$\&$g'$.
  \item Practically, the bluest bottom in the \(T_{\mathrm{eff}}\)-color diagram is not the real intrinsic color. Due to to the photometric uncertainty, the observed color will be either bluer or redder than its true color by the amount of photometric error, which leads to a distribution of color index on the bluer side of intrinsic color. Therefore, we take some percentage (X\%) of the bluest colors as the non-reddening region of each bin.
      \cite{jian_revision_2017} discussed the difference among choosing 3\%, 5\% and 10\% bluest stars, and concluded that the differences of color indexes are within 0.02 magnitude for A-M type stars in 2MASS/$J$ - 2MASS/$H$, which is smaller than the photometric error. Meanwhile, \citet{wang_universality_2014}, \citet{xue_precise_2016}, and \citet{jian_revision_2017} all chose 5\%. We take a slightly different strategy. For \(T_{\mathrm{eff}}\) between 8000\,K to 10000\,K, we follow the convention, i.e. choosing the 5\% bluest. While for \(T_{\mathrm{eff}}\) $>$ 10000\,K, 1\% is chosen. This change is made to coincide with the selection effect. The OB stars become intrinsically bright with high  \(T_{\mathrm{eff}}\) (here $>$ 10000\,K). If they experience no interstellar extinction, which means they are nearby, they are apparently very bright and easily become saturated in photometry, in particular in the optical bands. In such case, they are removed by our quality control processing. Consequently, the remained sample would contain smaller portion of stars with zero-reddening. The choice of the boundary temperature is somehow arbitrary, but 10000\,K marks the borderline between A-type and OB-type stars and also between the two catalogs. It can be anticipated that this change would bring about some discontinuity of the median values of the colors around 10000\,K, but it is small and being smoothed very well by the analytic fitting (c.f. next Section and Figure~\ref{fig:ResultMain}).


  \item Clip the points with a 3-sigma rule after a Gaussian fitting is performed to each bin sample stars, afterwards the median of the remaining points is taken  as the intrinsic color of the corresponding bin.
  \item Fit the blue edge points by the function :
  $$
  C_{\lambda_{1}, \lambda_{2}}^{0} = A \cdot \exp {(-B \cdot \frac{T_{\mathrm{eff}}}{10000  \, \rm K } )}+C
  $$
  Although the form of the fitting function is not important, the exponential form is chosen since a monotonously decreasing relation of the color index  with \(T_{\mathrm{eff}}\)  is expected for a blackbody radiation that is a very good approximation for early-type stars.
\end{enumerate}

\begin{figure}
    \centering
    \includegraphics[width=0.9\textwidth]{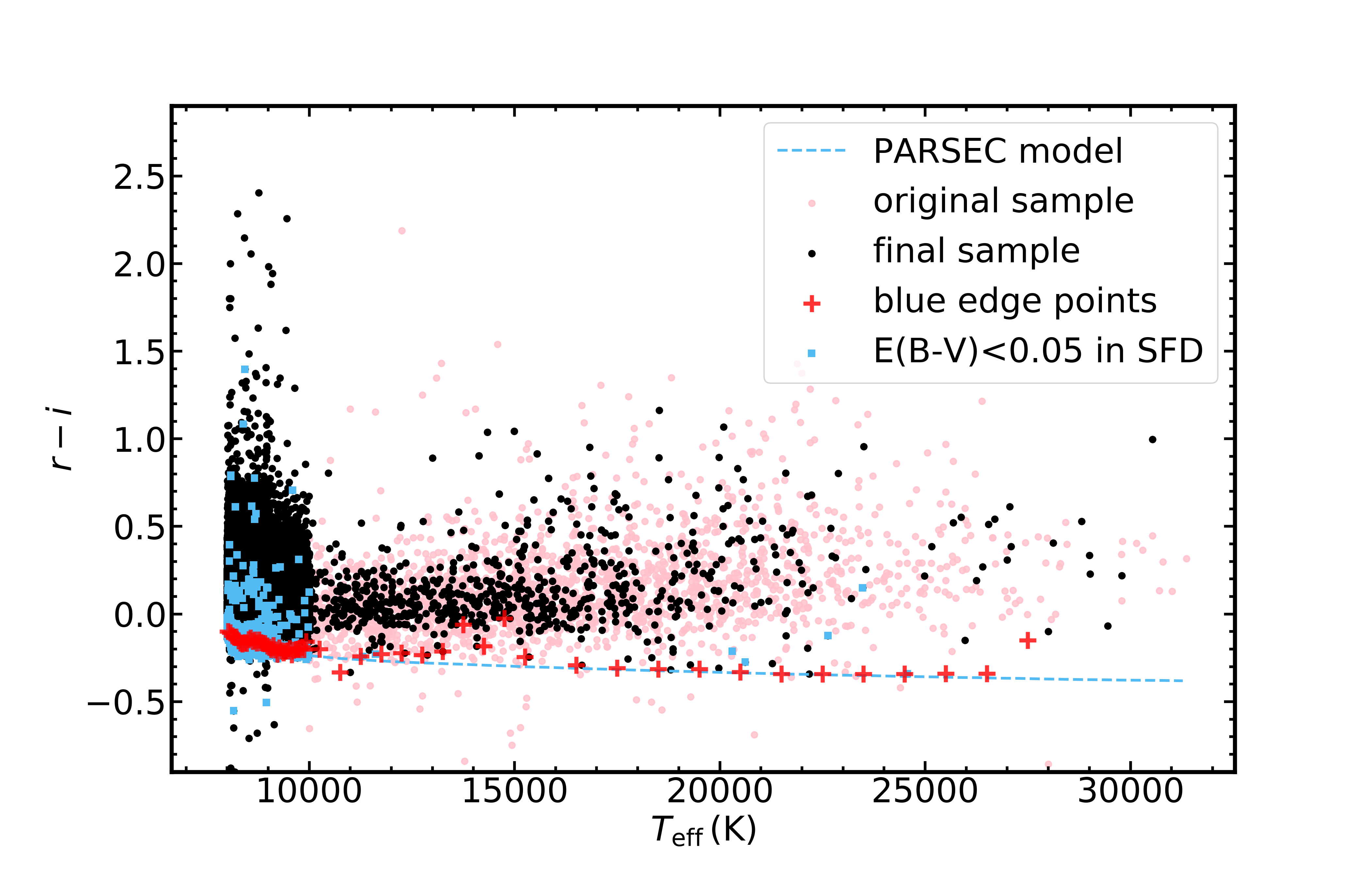}
    \caption{The color-\(T_{\mathrm{eff}}\) diagram for PS1/$r$ - PS1/$i$. The original sample is decoded by pink dots and the final sample by black dots. The median value of the bluest 5\% for \(T_{\mathrm{eff}}\) $<$ 10000\,K  and 1\% for \(T_{\mathrm{eff}}\) $>$ 10000\,K are denoted by red crosses. In comparison, the intrinsic colors calculated by the PARSEC model are displayed by blue dash line and the stars with $E(B-V)<0.05$ in the SFD dust map are displayed by blue squares.
    \label{fig:RatioExplain}}
\end{figure}

\begin{figure}
    \centering
    \includegraphics[width=0.85\textwidth]{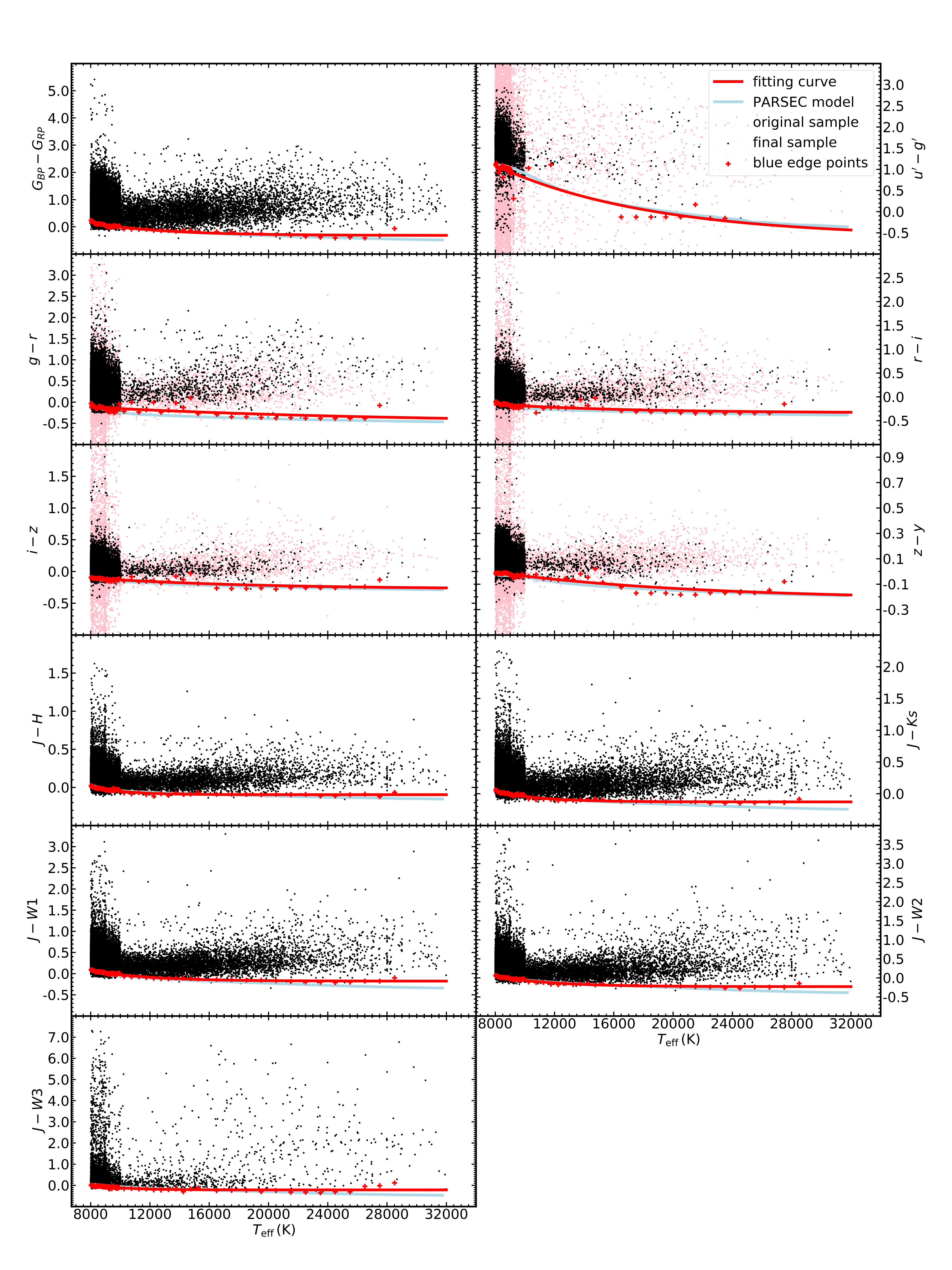}
    \caption{The color-\(T_{\mathrm{eff}}\) diagram for the studied color indexes. The original sample is decoded by pink dots and the final sample by black dots. The median value of the bluest 5\% for \(T_{\mathrm{eff}}\) $<$ 10000\,K  and 1\% for \(T_{\mathrm{eff}}\) $>$ 10000\,K are denoted by red crosses. In comparison, the intrinsic colors calculated by the PARSEC model are displayed by light blue line.
    \label{fig:ResultMain}}
\end{figure}

Figure~\ref{fig:RatioExplain} takes the color index PS1/$r$ - PS1/$i$ as an example to illustrate the method, where the black dots are the sample stars after quality control and the pink dots are from the original sample. It can be seen that many bluest stars with \(T_{\mathrm{eff}}\) $>$ 10000\,K are removed after quality control mostly due to saturation problem, which made us shift the percentage to 1\% when choosing the zero-reddening stars.
The median value of $r-i$ of the bluest 5\% (for \(T_{\mathrm{eff}}\) $<$ 10000\,K) or 1\% (for \(T_{\mathrm{eff}}\) $>$ 10000\,K) displays some deviations from the general tendency around 10000\,K and 16000\,K. The deviation around 10000\,K is caused by the change of the percentage, and that around 16000\,K is caused by the change of the bin width that a large bin width and a sliding window is taken for \(T_{\mathrm{eff}}\) $>$ 16000\,K to compensate for the reduction of the number of stars. But these deviations have little effect on the final fitting curve, which can be seen from Figure~\ref{fig:ResultMain}.
The relation of the intrinsic color index with \(T_{\mathrm{eff}}\) is compared with the PARSEC model, and a very good agreement is found both in Figure~\ref{fig:RatioExplain} and~\ref{fig:ResultMain}.
As \citet{sun_ultraviolet_2018} obtained the results by taking the very low-extinction stars from the SFD dust map \citep{schlegel_application_1998, schlegel_maps_1998, schlafly_measuring_2011} that is very consistent with the blue-edge method, we checked the stars with $E(B-V) < 0.05$\,mag in the SFD map.
It can be seen from Figure~\ref{fig:RatioExplain} that many of these stars suffer some amount of extinction and few OB stars are of low extinction, which can be understood that these early-type stars are located in a place further than the dust cloud traced by the SFD map, proving that they are distant. The method of taking the low-extinction stars from the SFD map works no more for the OB-type stars.

Figure~\ref{fig:ResultMain} displays the color-\(T_{\mathrm{eff}}\) diagrams for all the studied color indexes, where the symbols follow the convention of Figure~\ref{fig:RatioExplain}.
In addition, the fitting curves from the blue-edge method are present as well as the result from the PARSEC model calculation.

\begin{figure}
  \centering
  \includegraphics[width=0.7\textwidth]{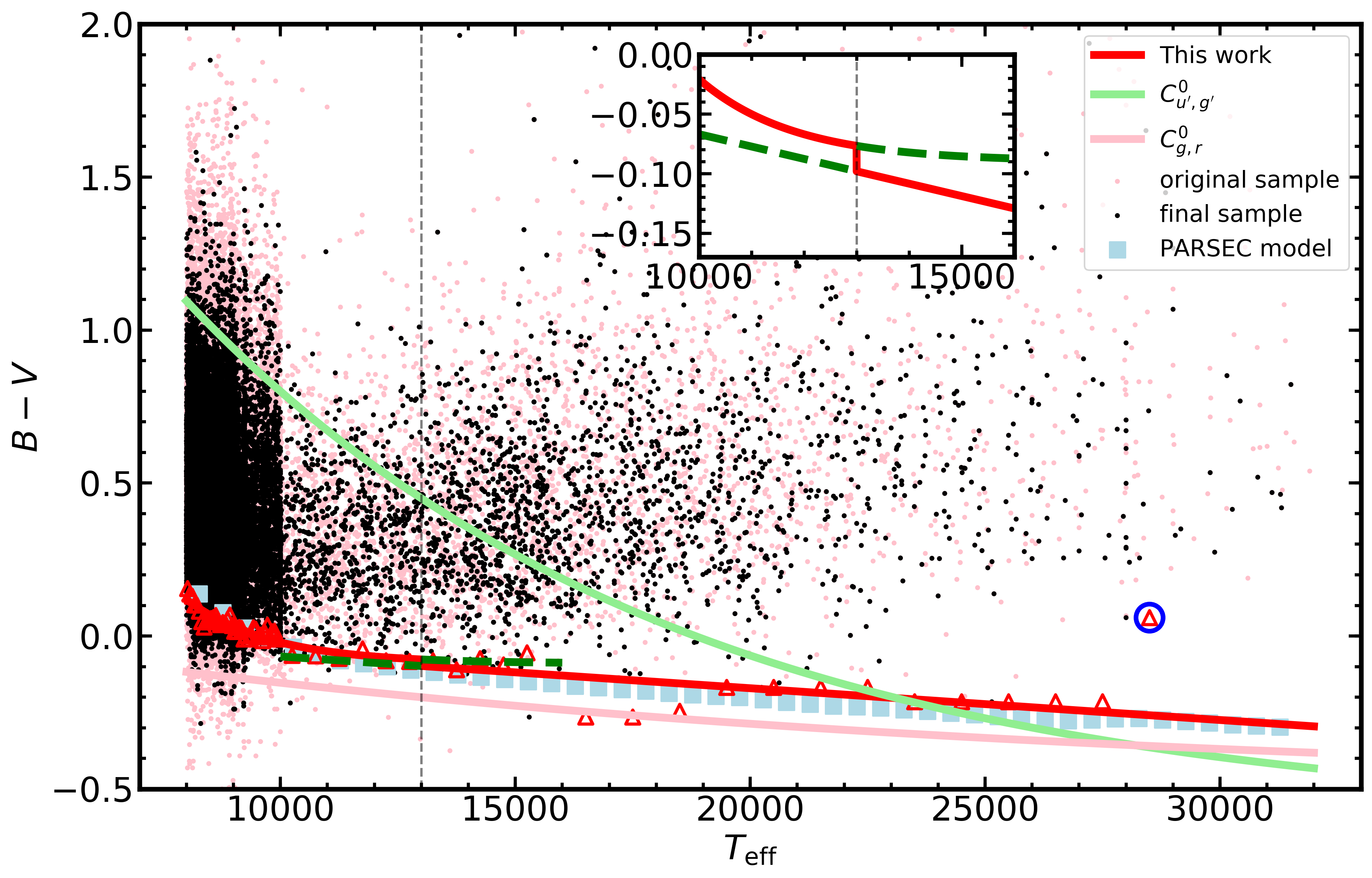}
  \caption{The $B-V$ vs \(T_{\mathrm{eff}}\) diagram. For \(T_{\mathrm{eff}}\) $>$ 16000\,K, the median value of the bluest 0.1\% is chosen to be the blue edge points, which also represent the bluest points in each bin. The exponential fit curve and linear fit line are adopted before and after 13000\,K, respectively. For the intrinsic color at \(T_{\mathrm{eff}}\) $=$ 13000\,K, the average of these two fitting curves is adopted. The intrinsic colors of $u'-g'$ and $g-r$ of this work are also presents. The red point at the  highest temperature with a blue circle is not included in the fitting. The inset displays the common part used in the two functions where the green dashed line denotes the rejected. The other symbols follow the convention of Figure~\ref{fig:ResultMain}.
  \label{fig:ResultBV}}
\end{figure}

The color index $B-V$ is dealt with in a way slightly different from the other color indexes. The blue edge points selected by the above criteria are apparently redder than the expectation from the PARSEC model, which is remarkable. After all, the color index $u'-g'$ should be more susceptible to extinction than $B-V$, but it shows high consistency with the PARSEC model. The color index $g-r$ very similar to $B-V$ neither deviates from PARSEC model. As discussed above, it may be caused by the exclusion of high temperature, zero-reddening stars due to the saturation problem.
To testify this possibility, all the APASS stars with \(T_{\mathrm{eff}}\) $>$ 10000\,K are included, i.e. no quality control is applied except those without error assignment. The newly determined curve (red solid line in Figure~\ref{fig:ResultBV}) is consistent with the PARSEC model, with a goodness comparable to other visual color indexes. This result may imply that the photometry quality control was too severe for the APASS catalog.
Another problem with $B-V$ is that a single exponential function cannot match the selected blue edge points well. In order to reflect the true trend, a linear function ($C_{B, V}^{0} = A \cdot \frac{T_{\mathrm{eff}}}{10000  \, \rm K } + B$) is introduced. Specifically, an exponential function is fitted for the points with \(T_{\mathrm{eff}}\) $<$ 16000\,K, and a linear function is fitted for the points with \(T_{\mathrm{eff}}\) $>$  10000\,K.
The points with \(T_{\mathrm{eff}}\) $\in [10000, 16000]$ are adopted to fit both the exponential and linear functions, which approximate each other closest at about 13000\,K, with a difference of $\simeq$ 0.02\,mag. So the final result takes the exponential curve at \(T_{\mathrm{eff}}\) $<$ 13000\,K and the linear curve at \(T_{\mathrm{eff}}\) $>$ 13000\,K, and the average color of them at \(T_{\mathrm{eff}}\) $=$ 13000\,K is adopted. The comparison with $u'-g'$ and $g-r$ in Figure~\ref{fig:ResultBV} shows a reasonable consistency.

%

\begin{figure}
    \centering
    \includegraphics[width=0.7\textwidth]{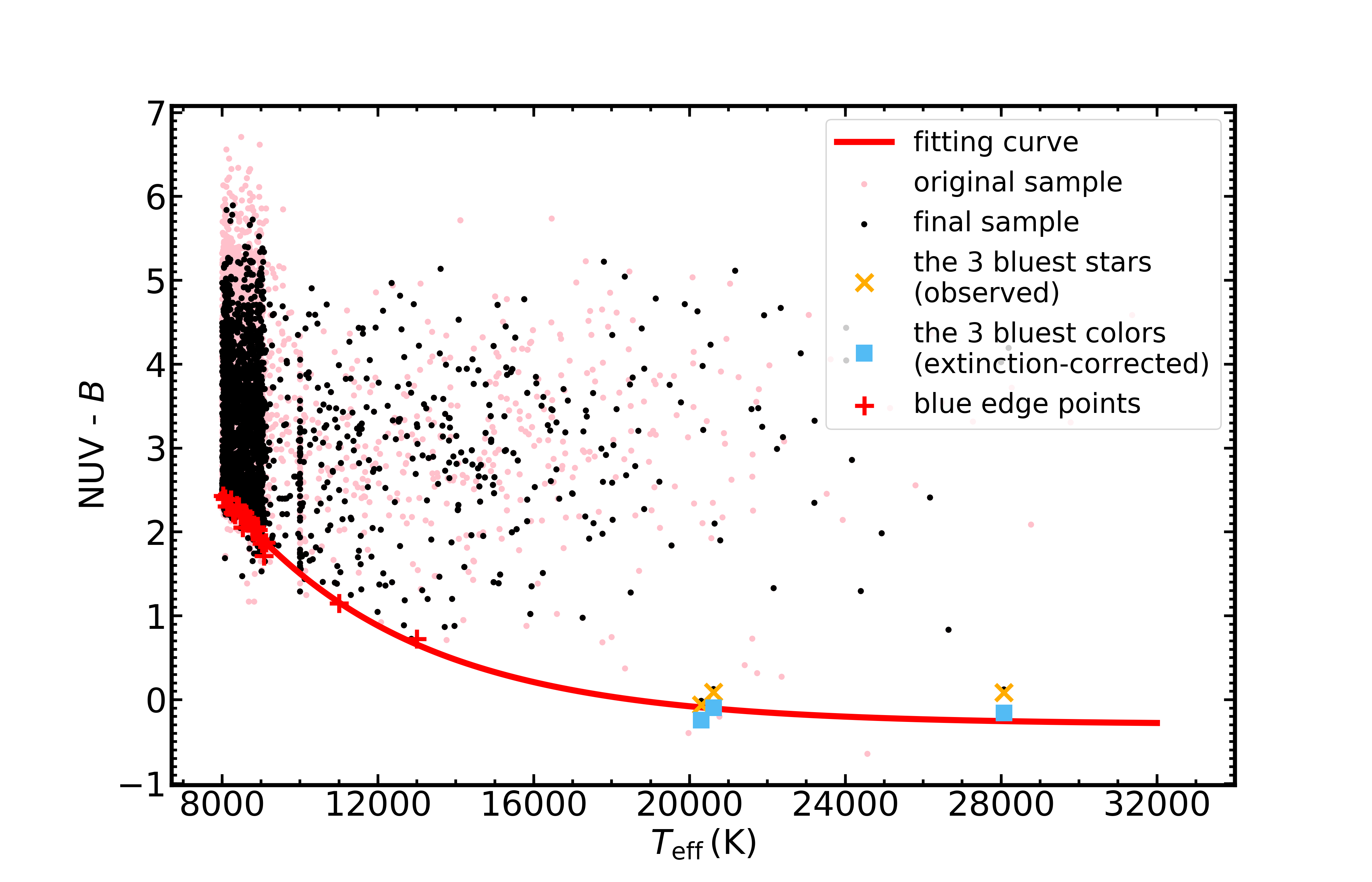}
    \caption{The $\mathrm{NUV}-B$ vs \(T_{\mathrm{eff}}\) diagram. The three bluest stars with \(T_{\mathrm{eff}}\) $>$ 16000\,K are chosen and corrected for interstellar extinction according to the $E(B-V)$ from the SFD dust map and the extinction law of \citet{sun_ultraviolet_2018}.  The other symbols follow the convention of Figure~\ref{fig:ResultMain}.
    \label{fig:Resultultraviolate}}
\end{figure}

The color index $\mathrm{NUV}-B$ is  treated specially and shown in Figure~\ref{fig:Resultultraviolate}.
As we mentioned above, a bin size of 2000\,K instead of 500\,K is adopted for $\mathrm{NUV}-B$ in the range of 10000\,K to 16000\,K. Moreover, the number of stars with \(T_{\mathrm{eff}}\) $>$ 16000\,K decreases sharply and there is no possibility to define some blue edge points there.
For a complete coverage of the temperature, we pick up the three bluest stars (denoted by cross in Figure~\ref{fig:Resultultraviolate}) and correct the interstellar extinction to obtain their intrinsic color indexes.
According to the SFD dust map, the $ E(B-V)$ towards their sightlines is $\sim$ 0.05 ($ E(B-V) \approx 0.05 $).
In addition, these 3 stars have high Galactic latitude that supports the correctness of the low extinctions.
From the average relation $ E(\mathrm{NUV}-\emph{B})/E(B-V) = 3.77$ by \citet{sun_ultraviolet_2018}, their intrinsic $\mathrm{NUV}-B$ is calculated.
Certainly the result depends on the accuracy of photometry and spectroscopy of the three stars and may suffer bigger uncertainty in comparison with other color indexes.
But the data is limited by the observation, this is the reliable result achievable up to date.

\section{Results and Discussions}
\label{sec:Results}

The results are presented in Figure~\ref{fig:ResultMain},~\ref{fig:ResultBV} and~\ref{fig:Resultultraviolate} to show the selection and fitting of the blue edge, and Table~\ref{tab:fittingparameter} for the fitting coefficients to the exponential function and the linear function for $C^0_{B,V}$ at \(T_{\mathrm{eff}}\) $>$ 13000\,K.
The intrinsic color indexes of spectral type O9-A5 are listed in Table~\ref{tab:ICtoST}.

For the  color indexes in the SDSS and Pan-STARRS1 bands, i.e. $u'-g'$, $g-r$, $r-i$, $i-z$, and $z-y$, the number of stars decreases apparently after quality control as shown in Figure~\ref{fig:ResultMain}. Like the case we have explained for $r-i$ in Figure~\ref{fig:RatioExplain}, the decreasing is caused by discarding the saturated stars that are nearby bright thus little reddened. Nevertheless, the blue edge is very consistent with the PARSEC model and the result should be reliable. It deserves to mention that the coincidence of $u'-g'$  with the PARSEC model is somehow accidental. Due to the small number of stars at high temperature and possibly the decreasing accuracy of photometry for bright stars,  it can be seen from Figure~\ref{fig:ResultMain} that the selected blue edge points are not numerous enough to trace the trend completely of the color change with \(T_{\mathrm{eff}}\) even with some relaxed constraint to the data. We would rather conclude that the result of $u'-g'$ is not in conflict with the PARSEC model. On the other hand, the $G_{BP}-G_{RP}$ is much less affected by the saturation problem.
The infrared bands have no problem of saturation either.


\subsection{Comparison with other results}\label{subsec:Comparison}

\begin{figure}
    \centering
    \includegraphics[width=0.8\textwidth]{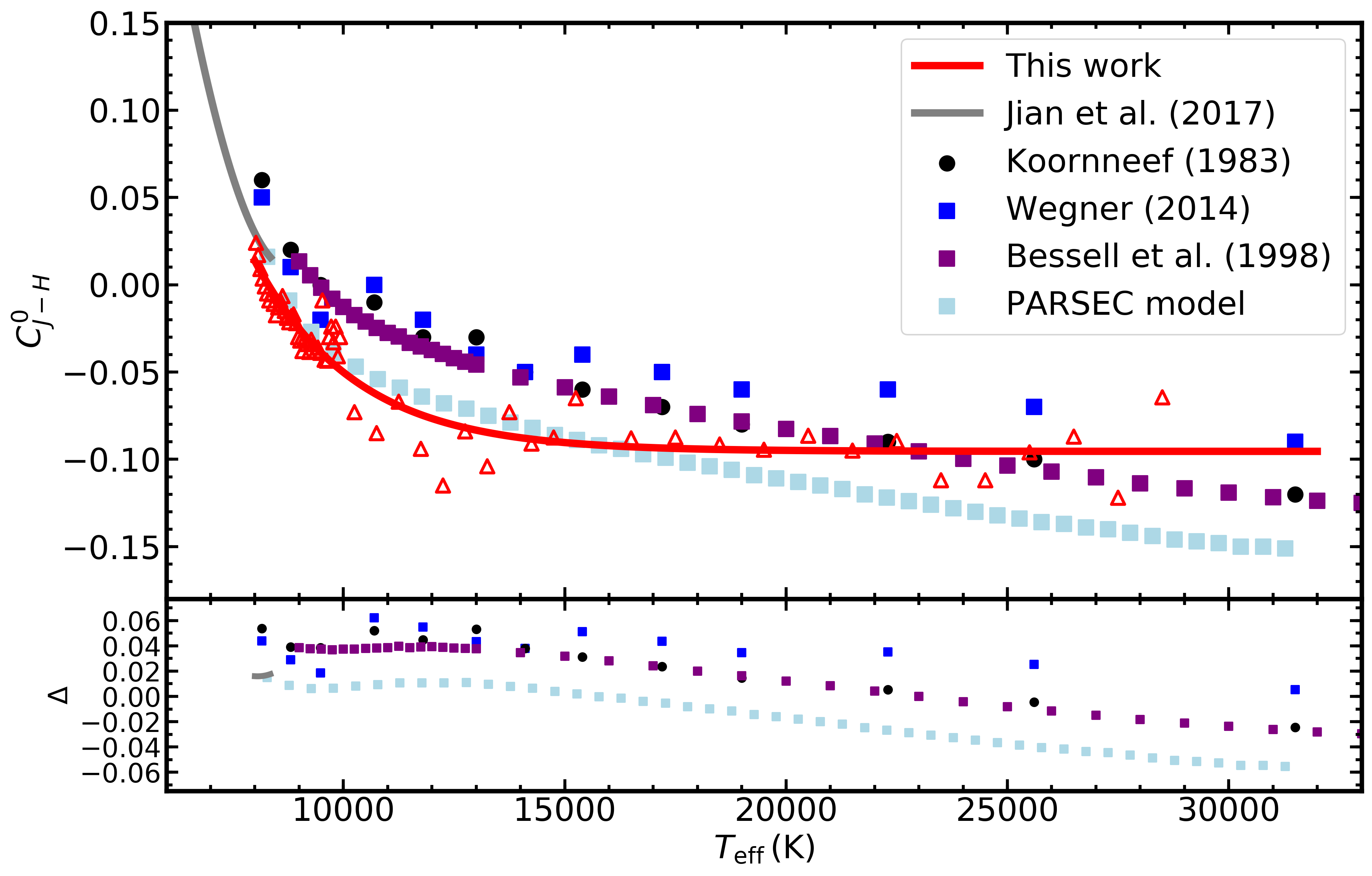}
    \caption{Comparison of the derived intrinsic color index $C^0_{J-H}$ with the previous works. The lower panel shows the difference $\Delta = C^0_{\text{PreviousWork}} - C^0_{\text{ThisWork}}$.
    \label{fig:JHcompare}}
\end{figure}

The result on $J-H$ is compared with previous works in Figure~\ref{fig:JHcompare}.
Since this is an extension to high temperature of the work by \citet{jian_revision_2017}, a smooth connection is expected.
It can be seen that the connection with \citet{jian_revision_2017} is pretty good, with a difference of about 0.02 mag at the joining point.
It should be noted that \citet{jian_revision_2017} used a third-order polynomial function to express the change of the color index with \(T_{\mathrm{eff}}\) while this work uses an exponential function, so that the two curves have different slopes and no extrapolation should be made to either side.
However, it is clear that our results show larger uncertainties at this joining point. Therefore, for those stars with \(T_{\mathrm{eff}}\) $\approx$ 8000\,K, such as A5-type stars (8160\,K according to \citealt{cox_allens_2002}), we suggest use the colors of blue edge points (decoded as red triangle in Figure~\ref{fig:JHcompare}) as the intrinsic colors. That is to say, the average colors of three blue edge points (at 8125K, 8175K and 8225K) are adopted to represent the intrinsic colors of A5-type stars (see Table~\ref{tab:ICtoST}).
On the other hand, the derived color index is slightly redder than that calculated from the PARSEC model.
But \citet{koornneef_near-infrared_1983}, \citet{bessell_model_1998} and \citet{wegner_intrinsic_2014} also obtained a redder $J-H$ than the PARSEC model, which bring about their agreement with our result in particular at high temperature end.
The other infrared color indexes, $J-Ks$, $J-W1$ and $J-W2$ are all slightly redder at the high temperature end than the PARSEC model.
It should be noted that \citet{bessell_model_1998} determined intrinsic colors for OB dwarfs by using the model atmosphere ATLAS9 convolved with the filter response functions with \(\log\,g\) of 4, 4.5 and 5, respectively. Since we selected dwarfs by \(\log\,g\) $>3.5$, the average color from the three surface gravities are compared in Figure~\ref{fig:JHcompare}, which is also based on the fact that the intrinsic colors with \(\log\,g\) $\in [4,5]$ are more or less the same. Apparently $J-H$ from \citet{bessell_model_1998} is redder than the PARSEC model by about 0.03 mag despite they used the same method.

\begin{figure}
    \centering
    \includegraphics[width=0.8\textwidth]{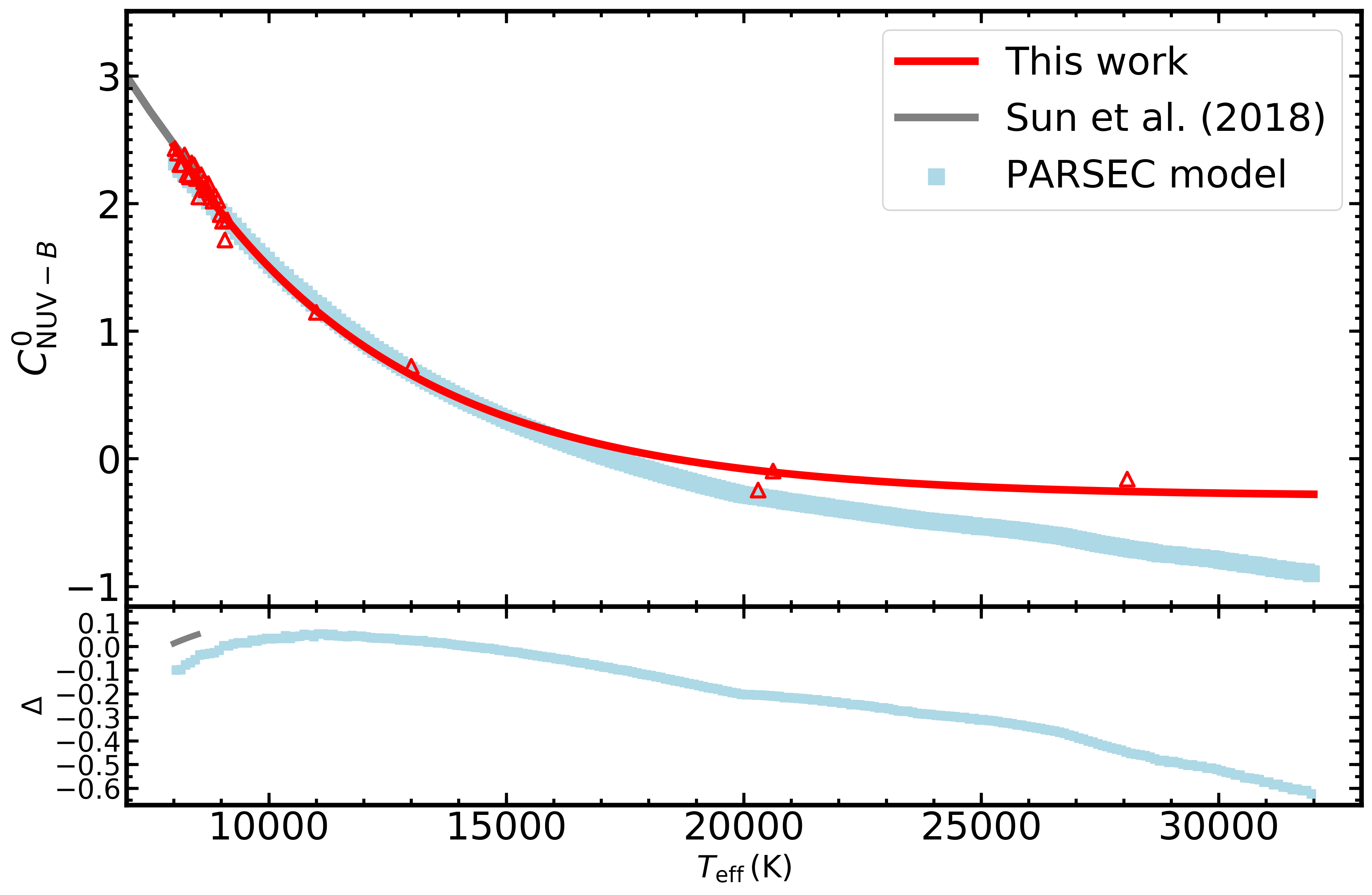}
    \caption{Comparison of the derived intrinsic color index $C^0_{\mathrm{NUV}-B}$ with \citet{sun_ultraviolet_2018} and the PARSEC model with $\mathrm{[Fe/H]} = 0$. The lower panel shows the difference $\Delta = C^0_{\text{PreviousWork}} - C^0_{\text{ThisWork}}$.
    \label{fig:NUVBcompare}}
\end{figure}

Similar to the case of $J-H$, the color index, $\mathrm{NUV}-B$, is an extension to the high temperature of the work of \citet{sun_ultraviolet_2018} that is compared in Figure~\ref{fig:NUVBcompare}.
Although $\mathrm{NUV}-B$ is sensitive to metallicity \citep{sun_ultraviolet_2018}, we make no division of metallicity because it is not measured for the LAMOST OB stars.
Fortunately OB stars are young massive stars and should be solar-like metal-rich.
So the case of $-0.125 < \mathrm{[Fe/H]} < 0.125$ from \citet{sun_ultraviolet_2018} is taken for comparison, the result is in good agreement.
Again, the color is redder than the PARSEC model at the high temperature end, like the case of $J-H$.
Unfortunately, no other work on $\mathrm{NUV}-B$ is available at this temperature range.
Since the blue edge is very difficult to define here, this discrepancy may be caused our over-estimation of the color index.
Meanwhile, the stellar model in the ultraviolet may suffer some uncertainty as well.

\begin{figure}
  \centering
  \includegraphics[width=0.8\textwidth]{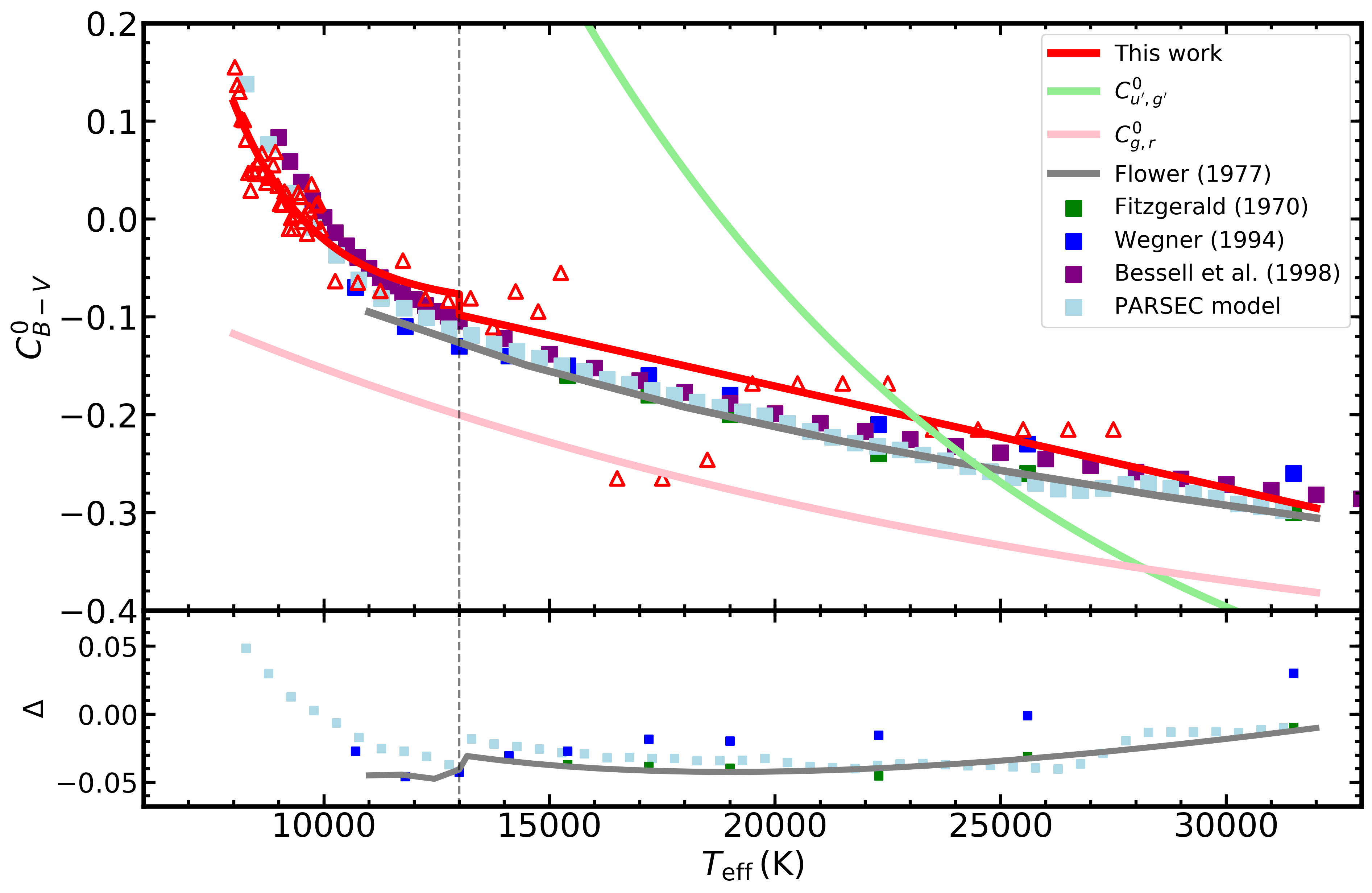}
  \caption{Comparison of the derived intrinsic color index $C^0_{B-V}$ with the previous works. The lower panel shows the difference $\Delta = C^0_{\text{PreviousWork}} - C^0_{\text{ThisWork}}$.
  \label{fig:BVcompare}}
\end{figure}

The color index $B-V$ agrees with previous work in general as shown in Figure~\ref{fig:BVcompare} in that the difference is mostly smaller than 0.05 mag. The coincidence is high with \citet{fitzgerald_intrinsic_1970}, \cite{flower_transformations_1977}, \cite{bessell_model_1998} and the PARSEC model. Meanwhile, the result of \citet{wegner_intrinsic_1994} seems to be redder than ours in particular at the high temperature end, which approaches our line of  $u'-g'$, nevertheless this difference is small and acceptable.

It should be mentioned that the optical color indexes are in very good agreement with the PARSEC model with no systematic deviations.

\subsection{The O-type stars}

\begin{figure}
\gridline{\fig{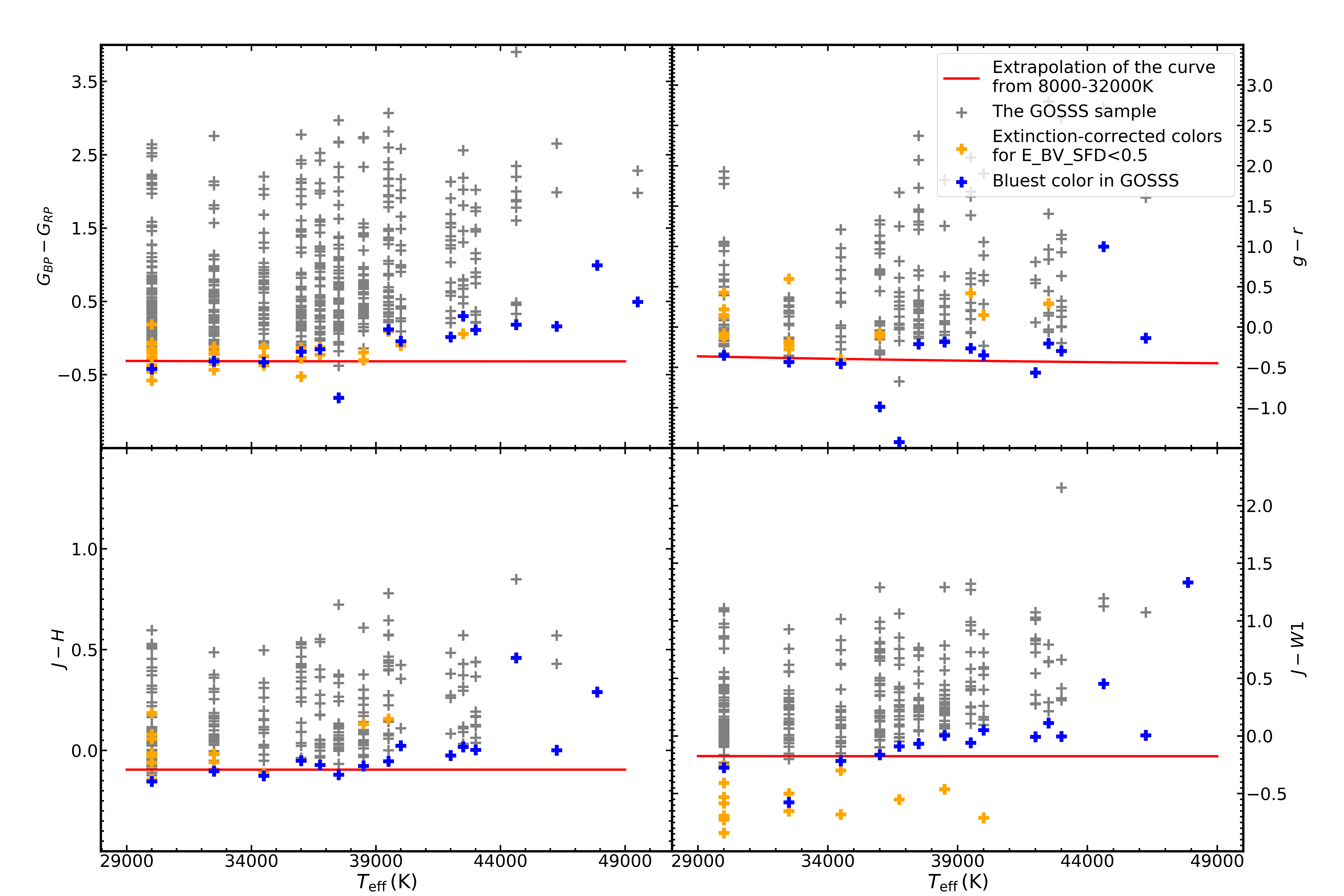}{0.9\textwidth}{}}
\caption{The color-\(T_{\mathrm{eff}}\) diagram for the O-type stars in the GOSSS catalog. The grey, blue and orange cross denote the GOSSS sample stars, the bluest star in each sub-type and the color index for stars with $E(B-V)<0.5$ after correcting for the interstellar extinction according to the $E(B-V)$ from the SFD dust map. The red line is the extrapolation from the fitting curves based on the sample of A- and B-type stars.
\label{fig:GOSSSestimation}}
\end{figure}

It can be seen from Table~\ref{tab:ICtoST} that the LAMOST catalogs contain mostly A- and B-type stars.
Concerning O-type stars, only very late-O type (O9) is included, the majority of O-type is missed.
For the completeness of this work, a sample of O-type stars is needed.
As mentioned in Section~\ref{sec:intro}, GOSSS is the largest catalog of pure O-type stars up to date that provide the spectral class but no \(T_{\mathrm{eff}}\) of the stars.
In order to use the GOSSS sample by the blue-edge method, the spectral type should be converted to the \(T_{\mathrm{eff}}\).
This is done by calculating the average \(T_{\mathrm{eff}}\) for each sub-type in \citet{massey_physical_2009}.
It should be mentioned that GOSSS determined the spectral types of some stars as O9.2 or O9.7 but the \(T_{\mathrm{eff}}\) for these two types are not available in \citet{massey_physical_2009} so that we adopted the rounding principle to approximate O9.2, O9.7 as O9, O9.5, respectively.
Furthermore, a few stars don't have accurate spectral type (marked as O4-5, etc.).
In this case, the former type (with higher \(T_{\mathrm{eff}}\)) is chosen to be its final type.
Figure~\ref{fig:GOSSSestimation} shows the results in four colors ($G_{BP}-G_{RP}$, $g-r$, $J-H$ and $J-W1$).
Since O-type stars are even more luminous than A- and B-type stars, they are mostly distant and suffer some interstellar extinction, thus there is little possibility to find zero-reddening stars in the sample.
Similar to what we have done for the $\mathrm{NUV}-B$  color, we checked the bluest colors and the colors after correcting for the extinction of the stars with $E(B-V) < 0.05$\,mag from the SFD dust map.
The conversion factors from $E(B-V)$ are $ E(G_{BP} - G_{RP})/E(B-V) = 1.321 $ and $ E(J - W1)/E(B-V) = 0.686 $ from \citet{wang_optical_2019}, and $ E(g-r)/E(B-V) = 1.018 $ and $ E(J-H)/E(B-V) = 0.260 $  from \citet{schlafly_measuring_2011}.
The locations of so derived ``intrinsic colors'' in Figure~\ref{fig:GOSSSestimation} obey no clear law, and the lack of such stars is evident at high temperature as expected.
Instead, the extrapolation of our analytic function of the relation between intrinsic color and \(T_{\mathrm{eff}}\) (red line in Figure~\ref{fig:GOSSSestimation}) seems to be able to delineate the tendency of the blue edge until the high edge of \(T_{\mathrm{eff}}\).
Therefore, we recommend the use of our analytic relations to the O-type stars even though the uncertainty should be borne in mind.

\subsection{The supergiants}

It has been well known in previous works that the intrinsic colors of supergiants and dwarfs differ significantly.
This work focuses on the OB dwarfs.
Since giants or supergiants are redder than dwarfs, the small amount of giants or supergiants that may be presented in the final sample after the data quality control would not change the intrinsic color indexes derived by the blue-edge method in this work.

The LAMOST sample includes supergiant stars in addition to dwarfs from which \citet{liu_catalog_2019} determined the luminosity class and spectral type for more than 8000 OB stars. The \citet{liu_catalog_2019} catalog contains no star in Class I or I-II, 208 stars in Class II, and 498 stars in Class II-III. We only take the 208 Class II stars as supergiants since Class III may be dwarf stars or mis-classified. The sample is enlarged by incorporating  the supergiant stars from the GOSSS catalog and \citet{hohle_masses_2010} with Class I, I-II and II. After cross-matching with 2MASS, there are 646 supergiants with observed $J-H$, specifically 89, 162 and 395 stars from LAMOST, GOSSS and \citet{hohle_masses_2010} respectively. For $B-V$, there are 450 supergiants, specifically 81, 112 and 257 stars from LAMOST, GOSSS and \citet{hohle_masses_2010} respectively. To include more supergiants in the final sample, the errors of $B$ and $V$ are required to be smaller than 0.1\,mag instead of 0.05\,mag for dwarfs; while the errors of $J$ and $H$ are still required to be less than 0.03\,mag.  Although the sample is insufficient to determine the intrinsic colors of supergiants by the blue-edge method, the previous results are testified with this new dataset.

\begin{figure}
  \gridline{
    \fig{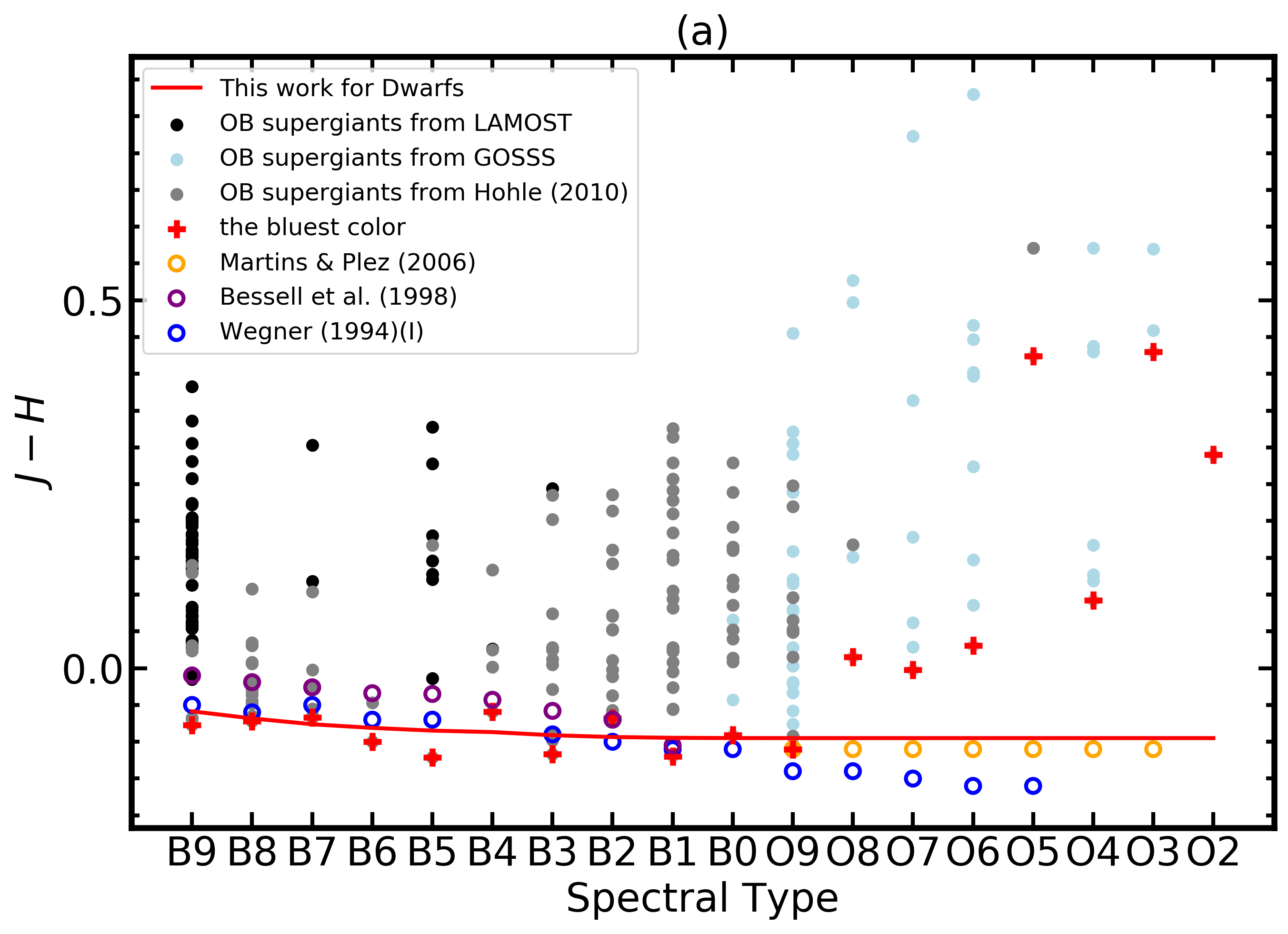}{0.5\textwidth}{}
    \fig{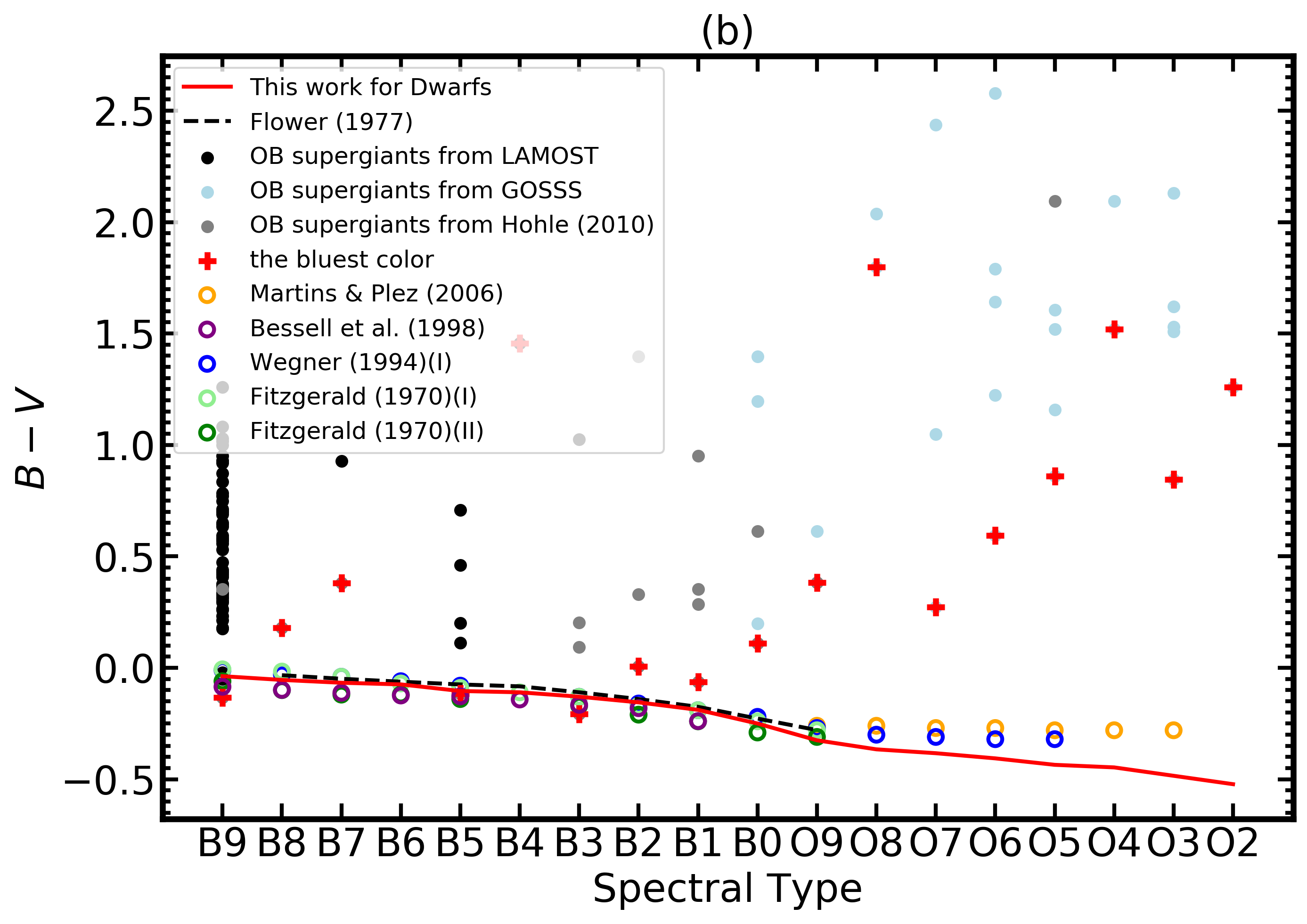}{0.515\textwidth}{}
            }
  \caption{The color-spectral type diagram for the supergiants (with luminosity class I, I-II and II). The black, light-blue and grey dots are the OB supergiants from LAMOST, GOSSS and \citet{hohle_masses_2010}, respectively. The red crosses denote the bluest color for each sub-type. Previous works are presented as circles. The red lines indicate intrinsic colors of dwarfs by this work.
  \label{fig:supergiants}}
\end{figure}

The distribution of the OB supergiants sample in the \(T_{\mathrm{eff}}\) vs. color index diagram is displayed in Figure~\ref{fig:supergiants} for  $J-H$ and $B-V$, where the bluest color in each sub-class is indicated by a red cross. Two obstacles prevent us from determining the intrinsic color by the blue-edge method. One is the very small sample which makes the statistical method invalid. The other is the lack of zero-reddening stars at high-temperature, the O-type stars. This can be understood because the OB supergiants are extremely luminous and mostly locate far in the Galactic plane with unavoidable significant interstellar extinction. Thus the colors are only compared with that of the dwarfs in this work and the supergiants by other works. The data of \citet{bessell_model_1998} are from the stars with \(\log\,g\) $=$ 2.5 and $=$ 3 at \(T_{\mathrm{eff}}\) $<$ 20000\,K and $>$ 20000\,K until 26000\,K respectively. In comparison with the dwarf stars (red curve in Figure~\ref{fig:supergiants}), it is difficult to draw any very certain conclusion from the data. However, the $(J-H)_0$ of supergiants derived from other works is apparently redder than that of B-type dwarfs. As for O-type stars, the color index $J-H$ of supergiants determined by \citet{wegner_intrinsic_1994} and \cite{martins_ubvjhk_2006} appear to be bluer than dwarfs. For $B-V$, the conclusion is different. While supergiants from other works appear very similar to dwarfs in this work for B-type stars, the results from \citet{wegner_intrinsic_1994} and \cite{martins_ubvjhk_2006} do appear redder than dwarfs for O-type stars which agree with previous conclusions.



\subsection{Uncertainty}\label{subsec:Uncertainty}

Generally the uncertainty of the intrinsic color indexes derived by the blue-edge method comes from three sources: (1) photometric uncertainty, (2) spectroscopic uncertainty, and (3) the uncertainty of fitting the relation of color index with \(T_{\mathrm{eff}}\).
As analyzed in \citet{jian_revision_2017}, source (3) mainly comes from how much percent of the bluest stars should be selected to represent the blue edge, which is usually less than the photometric error and negligible.
Then the uncertainty of the intrinsic color index $\varepsilon_{C_{\lambda_{1}, \lambda_{2}}}^0$  between band $\lambda_{1}$ and $\lambda_{2}$   can be written as:
$$
\varepsilon_{C_{\lambda_{1}, \lambda_{2}}}^0 =
\sqrt{
\varepsilon_{\lambda_{1}}^{2} +
\varepsilon_{\lambda_{2}}^{2} +
\varepsilon_{T_\mathrm{eff}}^{2}
}
$$
where $\varepsilon_{\lambda_{1}}$ and $\varepsilon_{\lambda_{2}}$ refer to the uncertainties from photometry in band $\lambda_{1}$ and $\lambda_{2}$  respectively, and $\varepsilon_{T_\mathrm{eff}}$ refers to the uncertainty brought by the error of \(T_{\mathrm{eff}}\).
$\varepsilon_{\lambda_{1}}$ and $\varepsilon_{\lambda_{2}}$ are chosen to be the upper limits of the photometric errors ($\sigma_{\lambda_{1}}$ and $\sigma_{\lambda_{2}}$ described in Section~\ref{sec:Data}).
As we choose 10 \% as the constraint of the relative error of temperature, the upper limit of \(T_{\mathrm{eff}}\) is about 1000\,K in the early-A type stars catalog.
Unfortunately, the OB stars catalog has no information on the error of \(T_{\mathrm{eff}}\).
According to the results of the intrinsic colors (Figure~\ref{fig:ResultMain}), the difference of 1000\,K in \(T_{\mathrm{eff}}\)  leads to a difference of about 0.02\,mag in all the color indexes except $\mathrm{NUV}-B$.
For $\mathrm{NUV}-B$,  this difference rises to about 0.2\,mag, which is taken to be the value of $\varepsilon_{T_\mathrm{eff}}$ for $\mathrm{NUV}-B$.
For other color indexes,  0.02 is used for $\varepsilon_{T_\mathrm{eff}}$.
The calculated uncertainty are presented in Table~\ref{tab:uncertainties}.
The error is $\sim$0.03\,mag in the optical, $\sim$0.05\,mag in the near-infrared, and $\sim$0.2\,mag in $\mathrm{NUV}-B$.

\section{Summary}
\label{sec:Summary}

By using the newly released largest photometric and spectroscopic catalogs for early-type stars, we determined their intrinsic color indices from the near-ultraviolet to infrared bands, namely \emph{Gaia}/$G_{BP}$,$G_{RP}$, SDSS/$u'$,$g'$, PS1/$g$,$r$,$i$,$z$,$y$, 2MASS/$J$,$H$,$K_S$, \emph{WISE}/$W1$,$W2$,$W3$, \emph{GALEX}/NUV and APASS/$B$,$V$.
The blue-edge method used results in an analytic relation of the intrinsic color index with the effective temperature \(T_{\mathrm{eff}}\) for B- and A-type stars. Our results agree well with the PARSEC model in the optical. Meanwhile, the derived intrinsic color indexes at the high temperature end are redder than the PARSEC model in the $B-V$, infrared and $\mathrm{NUV}-B$. The investigation of the GOSSS O-type stars seems to indicate that the derived relations of intrinsic colors with \(T_{\mathrm{eff}}\) can be extrapolated to O-type stars.
Our new data and results also support the intrinsic colors of B-type supergiants determined by previous works.

\acknowledgments Both Dingshan Deng and Yang Sun contributed equally to this paper. We are very grateful to  Profs. Wenyuan Cui and Chao Liu for their helpful discussions. We also thank the referee for his/her suggestions. This work is supported by the National Natural Science Foundation of China through the projects NSFC 11533002, 11603002 and Beijing Normal University grant No. 310232102.
This work has made use of data from the surveys by LAMOST, {\it Gaia}, SDSS, APASS, Pan-STARRS1, 2MASS, {\it WISE} and {\it GALEX}.
For SFD dust map \citep{schlegel_maps_1998}, we made use of the Python interface by \citet{green_dustmaps:_2018}.

\bibliography{OBIC}{}
\bibliographystyle{aasjournal}

\clearpage

\begin{deluxetable}{lccccccccccccc}
\tablecaption{The number of stars in the final sample for various color indexes \label{tab:samplenumber}}
\tablewidth{0pt}
\tablehead{
\colhead{Color index}  & \colhead{$C_{G_{BP},G_{RP}}$}      & \colhead{$C_{u',g'}$} & \colhead{$C_{g,r}$}  & \colhead{$C_{r,i}$}          & \colhead{$C_{i,z}$}  & \colhead{$C_{z,y}$}          & \colhead{$C_{J,H}$}  &
\colhead{$C_{H,Ks}$}  & \colhead{$C_{J,{W1}}$} & \colhead{$C_{J,{W2}}$} & \colhead{$C_{J,{W3}}$} & \colhead{$C_{B,V}$} & \colhead{$C_{{\rm NUV},B}$}
}
\startdata
    Total Numbers      & 140418  & 9391 & 55169   & 52071  & 50774        & 48682        & 67245        & 80158        & 99287        & 73655        & 7339 & 38760 &  2668        \\
\enddata
\end{deluxetable}

\begin{deluxetable}{llll}
\tablecaption{Coefficients of the function for fitting the relation of the intrinsic color with  \(T_{\mathrm{eff}}\) \label{tab:fittingparameter}}
\tablewidth{0pt}
\tablehead{
\colhead{Colors} & \colhead{A}   & \colhead{B}  & \colhead{C}
}
\startdata
$C^0_{G_{BP},G_{RP}}$ & 2.442  & 2.048 & -0.318 \\
$C^0_{u',g'}$ & 3.648	& 0.955	& -0.604 \\
$C^0_{g,r}$    & 0.566  & 0.486 & -0.501 \\
$C^0_{r,i}$    & 0.439  & 1.105 & -0.335 \\
$C^0_{i,z}$    & 0.362  & 0.906 & -0.278 \\
$C^0_{z,y}$    & 0.387  & 0.732 & -0.222 \\
$C^0_{J,H}$    & 3.662  & 4.392 & -0.095 \\
$C^0_{J,Ks}$    & 2.555  & 3.366 & -0.128 \\
$C^0_{J,{W1}}$  & 2.023  & 2.581 & -0.175 \\
$C^0_{J,{W2}}$ & 2.298  & 2.623 & -0.229 \\
$C^0_{J,{W3}}$ & 6.397  & 4.211 & -0.216 \\
$C^0_{B,V}$ \textbf{ (exponential)} & 17.546 & 5.539	& -0.090 \\
$C^0_{B,V}$ \textbf{ (linear)} & -0.116 & 0.051	&  \\
$C^0_{{\rm NUV},B}$   & 14.962 & 2.118 & -0.295
\enddata
\tablecomments{The first row for $C^0_{B,V}$ is the coefficients for exponential function ($C_{\lambda_{1}, \lambda_{2}}^{0} = A \cdot \exp {(-B \cdot \frac{T_{\mathrm{eff}}}{10000  \, \rm K } )}+C$) at \(T_{\mathrm{eff}}\) $<$ 13000\,K, while the second row is for linear function ($C_{B, V}^{0} = A \cdot \frac{T_{\mathrm{eff}}}{10000  \, \rm K } + B$) adopted at \(T_{\mathrm{eff}}\) $>$ 13000\,K. Other coefficients are all for the exponential function.}
\end{deluxetable}

\begin{deluxetable}{ccccccccccccccr}
\tablecaption{Intrinsic color indices according to spectral types \footnote{The relation between spectral type and \(T_{\mathrm{eff}}\) follows the Allen's definition \citep{cox_allens_2002}} \label{tab:ICtoST}}
\tablewidth{0pt}
\tablehead{
\colhead{Spectral Type}  & \colhead{$C^0_{G_{BP},G_{RP}}$}   & \colhead{$C_{u',g'}$}       & \colhead{$C^0_{g,r}$}  & \colhead{$C^0_{r,i}$}          & \colhead{$C^0_{i,z}$}  & \colhead{$C^0_{z,y}$}          & \colhead{$C^0_{J,H}$}  &
\colhead{$C^0_{J,Ks}$}  & \colhead{$C^0_{J,{W1}}$} & \colhead{$C^0_{J,{W2}}$} & \colhead{$C^0_{J,{W3}}$} & \colhead{$C_{B,V}$} & \colhead{$C^0_{{\rm NUV},B}$}
& \colhead{\(T_{\mathrm{eff}}\)}
}
\startdata
O9 & -0.32 & -0.44 & -0.38 & -0.32 & -0.26 & -0.19 & -0.09 & -0.13 & -0.17 & -0.23 & -0.22 & -0.33 & -0.28 & 32500 \\
B0 & -0.31 & -0.42 & -0.38 & -0.32 & -0.26 & -0.18 & -0.09 & -0.13 & -0.17 & -0.23 & -0.22 & -0.31 & -0.28 & 31500 \\
B1 & -0.31 & -0.29 & -0.34 & -0.31 & -0.24 & -0.16 & -0.09 & -0.13 & -0.17 & -0.23 & -0.22 & -0.25 & -0.23 & 25600 \\
B2 & -0.29 & -0.17 & -0.31 & -0.30 & -0.23 & -0.15 & -0.09 & -0.13 & -0.17 & -0.22 & -0.22 & -0.21 & -0.16 & 22300 \\
B3 & -0.27 & -0.01 & -0.28 & -0.28 & -0.21 & -0.13 & -0.09 & -0.12 & -0.16 & -0.21 & -0.21 & -0.17 & -0.03 & 19000 \\
B4 & -0.25 & 0.10  & -0.26 & -0.27 & -0.20 & -0.11 & -0.09 & -0.12 & -0.15 & -0.20 & -0.21 & -0.15 & 0.10  & 17200 \\
B5 & -0.21 & 0.23  & -0.23 & -0.25 & -0.19 & -0.10 & -0.09 & -0.11 & -0.14 & -0.19 & -0.21 & -0.13 & 0.28  & 15400 \\
B6 & -0.18 & 0.34  & -0.22 & -0.24 & -0.18 & -0.08 & -0.09 & -0.11 & -0.12 & -0.17 & -0.20 & -0.11 & 0.46  & 14100 \\
B7 & -0.15 & 0.45  & -0.20 & -0.23 & -0.17 & -0.07 & -0.08 & -0.10 & -0.10 & -0.15 & -0.19 & -0.08 & 0.66  & 13000 \\
B8 & -0.10 & 0.58  & -0.18 & -0.22 & -0.15 & -0.06 & -0.07 & -0.08 & -0.08 & -0.12 & -0.17 & -0.06 & 0.93  & 11800 \\
B9 & -0.05 & 0.71  & -0.16 & -0.20 & -0.14 & -0.05 & -0.06 & -0.06 & -0.05 & -0.09 & -0.15 & -0.04 & 1.26  & 10700 \\
A0 & 0.03  & 0.87  & -0.14 & -0.18 & -0.12 & -0.03 & -0.04 & -0.02 & 0.00  & -0.04 & -0.10 & 0.00  & 1.71  & 9480  \\
A2 & 0.08  & 0.97  & -0.13 & -0.17 & -0.12 & -0.02 & -0.02 & 0.00  & 0.03  & 0.00  & -0.06 & 0.04  & 2.02  & 8810  \\
A5 & 0.17  & 1.08  & -0.08 & -0.13 & -0.10 & -0.01 & 0.01 & 0.04  & 0.07 &  0.04 & -0.01 & 0.12 & 2.36  & 8160
\enddata
\tablecomments{As suggested in Section \ref{subsec:Comparison}, the intrinsic colors for A5-type stars are from the blue edge points instead of the fitting curve.}
\end{deluxetable}


\begin{deluxetable}{lccccccccccccc}
\tablecaption{Uncertainties of intrinsic color indices\label{tab:uncertainties}}
\tablewidth{0pt}
\tablehead{
\colhead{Uncertainties}  & \colhead{$C_{G_{BP},G_{RP}}$}     & \colhead{$C_{u',g'}$} & \colhead{$C_{g,r}$}  & \colhead{$C_{r,i}$}          & \colhead{$C_{i,z}$}  & \colhead{$C_{z,y}$}          & \colhead{$C_{J,H}$}  &
\colhead{$C_{H,Ks}$}  & \colhead{$C_{J,{W1}}$} & \colhead{$C_{J,{W2}}$} & \colhead{$C_{J,{W3}}$} & \colhead{$C_{B,V}$} & \colhead{$C_{{\rm NUV},B}$}
}
\startdata
$\varepsilon_{\lambda_{1}}$      & 0.01      & 0.05     & 0.02        & 0.02        & 0.02        & 0.02        & 0.03        & 0.03        & 0.03        & 0.03     & 0.03    & 0.05  & 0.10         \\
$\varepsilon_{\lambda_{2}}$      & 0.01      & 0.05     & 0.02        & 0.02        & 0.02        & 0.02        & 0.03        & 0.03        & 0.03        & 0.03     & 0.05    & 0.05  & 0.05        \\
$\varepsilon_{T_\mathrm{eff}}$   & 0.02      & 0.02     & 0.02        & 0.02        & 0.02        & 0.02        & 0.02        & 0.02        & 0.02        & 0.02     & 0.02    & 0.02  & 0.20         \\
$\varepsilon_{C_{\lambda_{1}, \lambda_{2}}}^0$ & 0.024  & 0.073 & 0.035 & 0.035    & 0.035       & 0.035       & 0.047       & 0.047       & 0.047       & 0.047    & 0.062   & 0.073 & 0.229 \\
\enddata
\end{deluxetable}

\clearpage

\end{document}